\documentclass[prl,superscriptaddress,twocolumn]{revtex4-2}
\usepackage{graphics,graphicx, array, afterpage}
\usepackage{qcircuit,amsmath}
\usepackage{grffile}
\usepackage[export]{adjustbox}
\usepackage{lineno}
\usepackage{xcolor}
\usepackage{longtable}
\usepackage{braket}
\usepackage{bbm}
\usepackage{ulem}
\usepackage{array}
\usepackage{multirow}
\usepackage{enumerate}
\usepackage{amsmath}
\usepackage{amssymb}
\usepackage{esint}
\usepackage{amsthm}
\usepackage{times}
\usepackage{tikz}
\usepackage{multirow}
\usepackage[colorlinks, citecolor=blue]{hyperref}
\hypersetup{linkcolor=magenta,urlcolor=blue,citecolor=blue,pdfstartview={FitH},urlcolor=blue}

\tikzset{every picture/.style={line width=0.75pt}} 

\begin{document}
\title{Exhaustive Construction of Generalized Parent Hamiltonians for Matrix Product States}

\author{He-Ran Wang}
\email{heranwang7@gmail.com}
\affiliation{Chen-Ning Yang Institute for Advanced Study, Tsinghua University, Beijing 100084,
People's Republic of China}

\author{Lorenzo Gotta}
\email{lorenzo.gotta@unige.ch}
\affiliation{Department of Quantum Matter Physics, University of Geneva, 1211 Geneva, Switzerland}

\author{Sanjay Moudgalya}
\email{sanjay.moudgalya@gmail.com}
\affiliation{Department of Theoretical Physics, Tata Institute of Fundamental Research, Mumbai 400005, India}
\affiliation{Technical University of Munich, TUM School of Natural Sciences, Physics Department, 85748 Garching, Germany}
\affiliation{Munich Center for Quantum Science and Technology (MCQST), Schellingstr. 4, 80799 M{\"u}nchen, Germany}

\begin{abstract}
Characterizing local Hamiltonians that host interesting many-body states as eigenstates has been an important problem in quantum many-body physics.
For matrix product states (MPS), there is a well-known construction of parent Hamiltonians that have the MPS as the frustration-free ground state.
However, recent results on quantum many-body scars have shown that such a construction does not exhaust all possible Hamiltonians that have the MPS as an eigenstate.
In this work, we develop an exhaustive construction of generalized parent Hamiltonians that have a given translation-invariant MPS as an eigenstate.
These go beyond the conventional projector-based construction of frustration-free parent Hamiltonians, where each local Hermitian term has the MPS as an eigenstate, by allowing non-Hermitian telescoping terms that cancel out with periodic boundary conditions (PBC). 
We derive a local tensor equation that establishes a correspondence between inequivalent classes of generalized parent Hamiltonians, i.e., those that cannot be related using linear combinations with frustration-free Hamiltonians, and the MPS transfer matrix spectrum.
We illustrate the connection to intrinsic physical properties, including non-Abelian symmetries and zero correlation length, and further extend this construction to block-injective MPS.
Our framework hence provides a systematic bridge between generalized parent Hamiltonians and the local structure of quantum states.
\end{abstract}

\maketitle

\textit{Introduction}---
One of the central problems in quantum many-body physics is to understand the structure of local Hamiltonians that realize interesting quantum states.
In this context, the framework of parent Hamiltonians starts from a desired quantum state and constructs local interactions for which that state is a ground state.
This perspective has played a foundational role in the development of exactly solvable models and in the understanding of quantum phases of matter, with paradigmatic examples ranging from valence bond states~\cite{affleck1987rigorous}, quantum spin liquids~\cite{Schroeter2007Spin, Thomale2009Parent} and quantum Hall systems~\cite{Laughlin1983Anomalous, Trugman1985Exact, Kapit2010Exact} to topological order~\cite{Kitaev2003Faulttolerant, Levin2005String} and tensor network states~\cite{Fannes1992Finitely, Perez2007Matrix, Perez2008PEPS, Cirac2021Matrix}.
More broadly, parent Hamiltonians provide a systematic bridge between the structure of a many-body wavefunction and the microscopic local interactions supporting it. They have therefore become increasingly relevant to quantum information tasks, including Hamiltonian learning~\cite{Chertkov2018Computational, Rattacaso2024Parent}, quantum simulation~\cite{Nascimbene2012Experimental, Semeghini2021Probing}, and quantum-state preparation~\cite{Diehl2008quantum, Verstraete2009Quantum, Schwarz2013Preparing}.

More recently, this perspective has been substantially generalized with the study of quantum many-body scars (QMBS)~\cite{Serbyn2021quantum, Moudgalya2022Quantum, Chandran2022Quantum}. QMBS are atypical nonthermal eigenstates embedded within an otherwise thermal spectrum of a local non-integrable Hamiltonian, thereby evading the Eigenstate Thermalization Hypothesis~\cite{Srednicki1994Chaos, Deutsch1991Quantum} that is generically expected of such Hamiltonians. 
The discovery of QMBS has broadened parent Hamiltonian constructions beyond ground states, motivating the systematic design of local Hamiltonians that host a given quantum state, or a set of states, as exact excited eigenstate(s) in their spectra.
Such Hamiltonians are referred to as generalized parent Hamiltonians.
Several formalisms have since been developed for obtaining examples of generalized parent Hamiltonians for QMBS~\cite{Shiraishi2017Systematic, Qi2019Determining, Moudgalya2020eta, Moudgalya2020Large, mark2020unified, Mark2020eta, shibata2020onsager,   nicholas2020from, Pakrouski2020Many, Pakrouski2021Group, Ren2021Quasisymmetry, ren2022deformed, Moudgalya2022Exhaustive, sanada2023quantum, moudgalya2023numerical, rozon2024broken, Gioia2026Distinct}.

Based on examples, previous studies pointed out an important distinction from the conventional parent Hamiltonian constructions: individual local Hamiltonian terms need not host the target states as eigenstates; it is sufficient that they do as a sum~\cite{lin2019exact, Mark2020eta, nicholas2020from, mark2020unified, Omiya2023Fractionalization, Wang2024Generalized, Ivanov2025Many, Gehrmann2026Exact, Rubio2026Local}.
This observation has motivated the three-fold classification of generalized parent Hamiltonians -- types I, II, and III-- that differ in the action of their local terms on the target states~\cite{Gioia2026Distinct}.
However, a systematic characterization of generalized parent Hamiltonians for general quantum states remains largely unexplored. Rigorous results have been developed only for certain target states~\cite{Ddea2026Locality, Omiya2026Local, Gioia2026Distinct}, while numerical search algorithms necessarily impose cutoffs on the ranges of local operators and the system sizes involved, hindering reliable extrapolations to the thermodynamic limit~\cite{Wang2024Generalized, Gioia2026Distinct}. 

In the present work, we address this problem for a broad class of matrix product states (MPS)~\cite{Perez2007Matrix, Cirac2021Matrix}.
MPS is an ansatz for area-law entangled quantum states and provides an ideal setting for initiating the study of generalized parent Hamiltonians.
MPS can efficiently represent many physically relevant states, including gapped ground states and short-time nonequilibrium states. 
Moreover, the local structure of an MPS is encoded in finite-dimensional tensors, which has recently enabled complete local characterizations of the global eigenstate condition~\cite{Rubio2026Local}.
These features make MPS particularly well suited for systematically studying how the local structure of the state constrains generalized parent Hamiltonians.

In the following, we first consider a single injective MPS, for which the three-fold classification of generalized parent Hamiltonians reduces to only types I and II, thanks to recent developments on their local characterization~\cite{Rubio2026Local}. 
Type I Hamiltonians can be constructed directly using local projectors, closely paralleling the conventional parent Hamiltonian construction.
Here we focus on the more nontrivial type II problem and develop a systematic method for exhaustively constructing all such Hamiltonians.
More specifically, we establish a linear algebraic equation in terms of the local MPS for constructing type II Hamiltonians, and derive an explicit counting of the number of inequivalent Hamiltonians, where equivalence is defined up to the addition of type I Hamiltonians~\cite{Moudgalya2022Exhaustive, Gioia2026Distinct, Supp}.
We then extend the construction to block-injective MPS, particularly including GHZ-type states.
Within this framework, we uncover systematic connections between type II Hamiltonians and intrinsic properties of the underlying MPS, including non-Abelian symmetries, correlation length, and spontaneous symmetry breaking, thereby unveiling a rich landscape of generalized parent Hamiltonians.

\textit{Type II Hamiltonians for MPS}---We begin by briefly reviewing the three-fold classification introduced in Ref.~\cite{Gioia2026Distinct}. Consider a set of quantum states $\{\ket{\psi_n}\}$ and a generalized parent Hamiltonian on a lattice, $H = \sum_j h_{[j]}$, where $h_{[j]}$ are strictly local \textit{Hermitian} terms with a finite range support around site $j$, and $H\ket{\psi_n} = E_n\ket{\psi_n}$. 
Here, the subscript $[j]$ denotes a support on a finite neighborhood of site $j$, while in a one-dimensional (1D) lattice the notation $j:j+r-1$ specifies the $r$ consecutive sites from $j$ to $j+r-1$.

For a type I Hamiltonian, there exists a Hermitian decomposition in which each individual local term has $\{\ket{\psi_n}\}$ as eigenstates. A type II Hamiltonian does not admit such a decomposition, but admits a non-Hermitian one: although $H$ itself is Hermitian, it can be expressed as a sum of strictly local \textit{non-Hermitian} terms, each of which has $\{\ket{\psi_n}\}$ as eigenstates. Hamiltonians that admit neither type of decomposition are classified as type III. 

In this work, we consider a homogeneous translation-invariant (TI) MPS on a 1D qudit chain with local Hilbert space dimension $q$ and length $L$ under PBC.
The wavefunction reads
\begin{equation}\label{eq:TI MPS}
 \ket{\psi(A)} = \sum_{a_1,\dots,a_L} \text{Tr}\left[A^{[a_1]}\cdots A^{[a_L]}\right] \ket{a_1, \dots,a_L},
\end{equation}
where $a_j = 1,2,\cdots,q$.
$\{A^{[a_j]}\}$ are a set of $\chi$-dimensional matrices acting on the virtual linear space, and $\chi$ is called the bond dimension. 
We focus on translation-invariant states, since the resulting constructions admit a natural extension to the thermodynamic limit, which is of primary physical interest.
We now present the graphical representation of an MPS:
\begin{equation*}
\tikzset{every picture/.style={line width=0.75pt}} 
\begin{tikzpicture}[x=0.75pt,y=0.75pt,yscale=-1,xscale=1]

\draw [fill={rgb, 255:red, 74; green, 144; blue, 226 }  ,fill opacity=1 ][line width=0.75]    (240.09,110.24) -- (240.09,131.02) ;
\draw [fill={rgb, 255:red, 74; green, 144; blue, 226 }  ,fill opacity=1 ][line width=0.75]    (142,110.23) -- (142,131) ;
\draw [line width=1.5]    (117.61,131.37) -- (195,131.23) ;
\draw  [fill={rgb, 255:red, 208; green, 2; blue, 27 }  ,fill opacity=1 ] (136,131) .. controls (136,127.69) and (138.69,125) .. (142,125) .. controls (145.31,125) and (148,127.69) .. (148,131) .. controls (148,134.31) and (145.31,137) .. (142,137) .. controls (138.69,137) and (136,134.31) .. (136,131) -- cycle ;
\draw [line width=1.5]  [dash pattern={on 1.69pt off 2.76pt}]  (195,131.23) -- (222.39,131.09) ;
\draw [fill={rgb, 255:red, 74; green, 144; blue, 226 }  ,fill opacity=1 ][line width=0.75]    (175.3,110.41) -- (175.3,131.18) ;
\draw  [fill={rgb, 255:red, 208; green, 2; blue, 27 }  ,fill opacity=1 ] (169.3,131.3) .. controls (169.3,127.98) and (171.99,125.3) .. (175.3,125.3) .. controls (178.62,125.3) and (181.3,127.98) .. (181.3,131.3) .. controls (181.3,134.61) and (178.62,137.3) .. (175.3,137.3) .. controls (171.99,137.3) and (169.3,134.61) .. (169.3,131.3) -- cycle ;
\draw [line width=1.5]    (222.39,131.09) -- (259.78,130.95) ;
\draw  [fill={rgb, 255:red, 208; green, 2; blue, 27 }  ,fill opacity=1 ] (234.3,131.3) .. controls (234.3,127.98) and (236.99,125.3) .. (240.3,125.3) .. controls (243.62,125.3) and (246.3,127.98) .. (246.3,131.3) .. controls (246.3,134.61) and (243.62,137.3) .. (240.3,137.3) .. controls (236.99,137.3) and (234.3,134.61) .. (234.3,131.3) -- cycle ;
\draw [line width=1.5]    (117.61,131.37) .. controls (27,172.23) and (368,170.23) .. (259.78,130.95) ;

\draw (133,138.4) node [anchor=north west][inner sep=0.75pt]    {$A$};
\draw (166,139.4) node [anchor=north west][inner sep=0.75pt]    {$A$};
\draw (232,138.4) node [anchor=north west][inner sep=0.75pt]    {$A$};
\draw (43,129.4) node [anchor=north west][inner sep=0.75pt]    {$\ket{\psi(A) } =$};

\end{tikzpicture}
\end{equation*}
where the open legs denote qudit degrees of freedom, and the contracted line represents a summation over bond indices.

First, we consider injective TI MPS~\cite{Fannes1992Finitely}.  Injectivity is characterized by an injectivity length $l_{\mathrm{inj}}$, defined as the minimum number of consecutive sites such that the products $\{\prod_{j=1}^{l_\text{inj}} A^{[a_j]} \}$ span the full algebra of $\chi\times\chi$ matrices.
The exhaustive construction of type I Hamiltonians can be formulated through the Shiraishi-Mori embedding~\cite{Shiraishi2017Systematic}: for an arbitrary local operator range $r$, one first considers the Hilbert subspace spanned by $r$-local MPS:
\begin{equation*}
\tikzset{every picture/.style={line width=0.75pt}} 
\begin{tikzpicture}[x=0.75pt,y=0.75pt,yscale=-1,xscale=1]

\draw [fill={rgb, 255:red, 74; green, 144; blue, 226 }  ,fill opacity=1 ][line width=0.75]    (340.3,114.02) -- (340.3,133.79) ;
\draw [fill={rgb, 255:red, 74; green, 144; blue, 226 }  ,fill opacity=1 ][line width=0.75]    (242.22,113) -- (242.22,133.77) ;
\draw [line width=1.5]    (217.83,134.14) -- (295.22,134) ;
\draw  [fill={rgb, 255:red, 208; green, 2; blue, 27 }  ,fill opacity=1 ] (236.22,133.77) .. controls (236.22,130.46) and (238.91,127.77) .. (242.22,127.77) .. controls (245.53,127.77) and (248.22,130.46) .. (248.22,133.77) .. controls (248.22,137.09) and (245.53,139.77) .. (242.22,139.77) .. controls (238.91,139.77) and (236.22,137.09) .. (236.22,133.77) -- cycle ;
\draw [line width=1.5]  [dash pattern={on 1.69pt off 2.76pt}]  (295.22,134) -- (322.61,133.86) ;
\draw [fill={rgb, 255:red, 74; green, 144; blue, 226 }  ,fill opacity=1 ][line width=0.75]    (275.52,113.18) -- (275.52,133.96) ;
\draw  [fill={rgb, 255:red, 208; green, 2; blue, 27 }  ,fill opacity=1 ] (269.52,134.07) .. controls (269.52,130.76) and (272.21,128.07) .. (275.52,128.07) .. controls (278.84,128.07) and (281.52,130.76) .. (281.52,134.07) .. controls (281.52,137.38) and (278.84,140.07) .. (275.52,140.07) .. controls (272.21,140.07) and (269.52,137.38) .. (269.52,134.07) -- cycle ;
\draw [line width=1.5]    (322.61,133.86) -- (360,133.72) ;
\draw  [fill={rgb, 255:red, 208; green, 2; blue, 27 }  ,fill opacity=1 ] (334.52,134.07) .. controls (334.52,130.76) and (337.21,128.07) .. (340.52,128.07) .. controls (343.84,128.07) and (346.52,130.76) .. (346.52,134.07) .. controls (346.52,137.38) and (343.84,140.07) .. (340.52,140.07) .. controls (337.21,140.07) and (334.52,137.38) .. (334.52,134.07) -- cycle ;
\draw  [line width=3] [line join = round][line cap = round] (234.14,109.55) .. controls (240.37,104.36) and (240.86,105.14) .. (251.52,105.41) .. controls (260.06,105.64) and (268.63,105.64) .. (277.17,105.41) .. controls (279.48,105.35) and (286.28,99.39) .. (286.28,96.45) .. controls (286.28,95.11) and (288.04,98.7) .. (288.76,99.9) .. controls (289.84,101.7) and (293.82,104.7) .. (295.38,104.72) .. controls (310.27,104.95) and (325.18,104.49) .. (340.07,104.72) .. controls (342.39,104.76) and (344.6,109.55) .. (345.86,109.55) ;

\draw (231.22,142.17) node [anchor=north west][inner sep=0.75pt]    {$A$};
\draw (267,142.4) node [anchor=north west][inner sep=0.75pt]    {$A$};
\draw (331.22,142.17) node [anchor=north west][inner sep=0.75pt]    {$A$};
\draw (149,114.4) node [anchor=north west][inner sep=0.75pt]    {$\ket{A_{\mu \nu }^{( r)}} =$};
\draw (204,127.4) node [anchor=north west][inner sep=0.75pt]    {$\mu $};
\draw (362,127.4) node [anchor=north west][inner sep=0.75pt]    {$\nu $};
\draw (272,74.4) node [anchor=north west][inner sep=0.75pt]    {$r\ \text{legs}$};

\end{tikzpicture}
\end{equation*}
where $\mu, \nu = 1,\cdots, \chi$ denote bond indices.
One then introduces the projector $Q^{(r)}$ onto the subspace spanned by $\{\ket{A^{(r)}_{\mu\nu}}\}$, and $P^{(r)}=I-Q^{(r)}$ for the orthogonal complement.
An $r$-local term of any type I Hamiltonian can then be constructed as $h_{j:j+r-1} = P^{(r)}_{j:j+r-1}D_{j:j+r-1}P^{(r)}_{j:j+r-1}$, where $D_{j:j+r-1}$ is an arbitrary Hermitian operator.
By construction, we have $h_{j:j+r-1}\ket{\psi(A)} = 0$, which realizes the type I condition~\cite{Gioia2026Distinct}.
Further choosing $D$ to be the identity makes $\ket{\psi(A)}$ a frustration-free ground state of $H$, thereby reducing to the conventional parent Hamiltonian construction~\cite{Fannes1992Finitely}.

For $r\ge l_\text{inj}$, injectivity implies $\operatorname{rank}{Q^{(r)}} = \chi^2$, and hence $\operatorname{rank}{P^{(r)}} = q^r -  \chi^2$, so the linear space of type I Hamiltonians grows exponentially with the operator range $r$.
As we show below, the type II Hamiltonians exhibit a similar growth, but the dimension is drastically reduced to a finite value after quotienting out the linear space of type I Hamiltonians, which defines the notion of a type II equivalence class~\cite{Moudgalya2022Exhaustive, Gioia2026Distinct}.
The result of recent work shows that any local term of a generalized parent Hamiltonian of an injective MPS must act as~\cite{Rubio2026Local}: 
\begin{equation}\label{eq:local_NH}
    h_{j:j+r-1}\ket{\psi(A)} = (\epsilon-X_{j:j+r-2}+X_{j+1:j+r-1})\ket{\psi(A)},
\end{equation}
where $X$ is in general non-Hermitian, and $\epsilon L = E$.
This relation applies for $r\ge l_\mathrm{inj}+1$, which can always be satisfied by grouping local Hermitian terms.
The key observation is that the local Hermitian term may fail to have $\ket{\psi(A)}$ as an eigenstate, but after adding a generally non-Hermitian telescoping term it should do so.
The modified local terms $(h_{j:j+r-1}+X_{j:j+r-2}-X_{j+1:j+r-1})$ therefore furnish a decomposition of $H$ as a sum of generally non-Hermitian local terms, which implies the absence of type III Hamiltonians~\cite{Gioia2026Distinct}.
Note that the operator $X$ need not be translation invariant a priori and can have explicit site-dependence, as the telescoping cancellation only requires invariance upon shifting around the whole system.
Nevertheless, for physical relevance, we focus on cases where the MPS is an eigenstate of the Hamiltonian for all sufficiently large system sizes $L$, which imposes periodicity on the site-dependence of $X$ and hence makes it translation invariant after grouping sufficiently many sites.
In the following, we provide an exhaustive characterization of type II Hamiltonians for a single injective TI MPS, thereby completing generalized parent Hamiltonian constructions in this setting.

\textit{The local tensor equations}---
We now introduce our exhaustive construction.
Since we consider a single state, we can w.l.o.g. set the eigenenergy to $E = 0$, and hence also $\epsilon = 0$.
We first fix the local operator range $r$ to construct the local terms of the Hamiltonian, and subsequently show that taking $r$ sufficiently large exhausts all possible equivalence classes of type II Hamiltonians.
We begin by rewriting Eq.~\eqref{eq:local_NH} in a local form for sufficiently large $L$,
\begin{equation}\label{eq:local_NH_ket}
    (h_{1:r}+X_{1:r-1}-X_{2:r})\ket{A_{\mu \nu }^{( r)}} = 0,
\end{equation}
where $\mu,\nu$ can be arbitrary.
Taking the Hermitian conjugate gives
\begin{equation}\label{eq:local_NH_bra}
    \bra{A_{\mu' \nu' }^{( r)}}(h_{1:r}-X_{1:r-1}+X_{2:r}) = 0,
\end{equation}
where we have taken $X$ to be anti-Hermitian w.l.o.g. by absorbing the Hermitian part into $h$.
We then contract Eq.~\eqref{eq:local_NH_ket} with $\bra{A_{\mu' \nu' }^{( r)}}$ and Eq.~\eqref{eq:local_NH_bra} with $\ket{A_{\mu \nu }^{( r)}}$. Subtracting the two resulting expressions eliminates $h$ and gives
\begin{equation}\label{eq:transfer_matrix_formula}
    \bra{A_{\mu' \nu' }^{( r)}}X_{1:r-1}\ket{A_{\mu \nu }^{( r)}} = \bra{A_{\mu' \nu' }^{( r)}}X_{2:r}\ket{A_{\mu \nu }^{( r)}}.
\end{equation}

Given the local tensor $A$, Eq.~\eqref{eq:transfer_matrix_formula} provides a set of homogeneous linear equations for the operator $X$. 
Since the global Hamiltonian is the object of physical interest, it remains to construct a local Hermitian operator $h$ from the solution $X$.
Using the $r$-local projectors introduced above, we choose
\begin{align}\label{eq:reconstruction}
    h_{1:r} = &Q^{(r)} (X_{1:r-1} - X_{2:r}) P^{(r)}\nonumber\\
    &- P^{(r)} (X_{1:r-1} - X_{2:r}) Q^{(r)}\nonumber\\
    =&[Q^{(r)}, X_{1:r-1} - X_{2:r}].
\end{align}
Since $X$ is anti-Hermitian, we have $h_{1:r} = h_{1:r}^\dagger$, as well as the local annihilation condition that follows from Eq.~(\ref{eq:local_NH_ket}):
\begin{align}
&(h_{1:r}+X_{1:r-1}-X_{2:r})Q^{(r)}
\nonumber\\
=&(I-P^{(r)})(X_{1:r-1}-X_{2:r})Q^{(r)}
\nonumber\\
=&Q^{(r)}(X_{1:r-1}-X_{2:r})Q^{(r)}=0.
\end{align}
The construction above does not introduce any explicit site-dependence on the $X$ operator, which, as we discussed earlier, is sufficient for the analysis of TI MPS.
Moreover, even though generalized parent Hamiltonians can possess site-dependent type I terms, they do not change the equivalence class of the type II Hamiltonian. Since here we are only interested in such classes, we can choose a translation-invariant representative.

\textit{Characterizing the solution space}---Solutions of Eq.~\eqref{eq:transfer_matrix_formula} form a linear space of local operators.
Through the construction~\eqref{eq:reconstruction}, we can obtain a generalized parent Hamiltonians.
However, distinct solutions for $X$ may correspond to type II Hamiltonians equivalent up to the addition of type I Hamiltonians, and should be regarded as redundant for the characterization of equivalence classes of generalized parent Hamiltonians.
Below, we formulate Eq.~\eqref{eq:transfer_matrix_formula} in terms of the MPS transfer matrix, and characterize the redundancy and the number of inequivalent type II Hamiltonians in terms of the commutant of the transfer matrix.
To this end, we first present the graphical form of Eq.~\eqref{eq:transfer_matrix_formula} (taking $r=3$ for illustration):
\begin{equation}\label{eq:transfer_matrix_graph}
\tikzset{every picture/.style={line width=0.75pt}} 
\begin{tikzpicture}[x=0.75pt,y=0.75pt,yscale=-1,xscale=1]

\draw [fill={rgb, 255:red, 74; green, 144; blue, 226 }  ,fill opacity=1 ][line width=0.75]    (352.3,130.3) -- (352.3,140.07) ;
\draw [fill={rgb, 255:red, 74; green, 144; blue, 226 }  ,fill opacity=1 ][line width=0.75]    (215.09,97.02) -- (215.09,140.79) ;
\draw [fill={rgb, 255:red, 74; green, 144; blue, 226 }  ,fill opacity=1 ][line width=0.75]    (184.3,96.07) -- (184.3,107.84) ;
\draw [fill={rgb, 255:red, 74; green, 144; blue, 226 }  ,fill opacity=1 ][line width=0.75]    (156,95.77) -- (156,108.55) ;
\draw [fill={rgb, 255:red, 74; green, 144; blue, 226 }  ,fill opacity=1 ][line width=0.75]    (156,130) -- (156,144.77) ;
\draw [line width=1.5]    (131.61,141.14) -- (240,141) ;
\draw  [fill={rgb, 255:red, 208; green, 2; blue, 27 }  ,fill opacity=1 ] (150,140.77) .. controls (150,137.46) and (152.69,134.77) .. (156,134.77) .. controls (159.31,134.77) and (162,137.46) .. (162,140.77) .. controls (162,144.09) and (159.31,146.77) .. (156,146.77) .. controls (152.69,146.77) and (150,144.09) .. (150,140.77) -- cycle ;
\draw [fill={rgb, 255:red, 74; green, 144; blue, 226 }  ,fill opacity=1 ][line width=0.75]    (184.3,130.18) -- (184.3,140.96) ;
\draw  [fill={rgb, 255:red, 208; green, 2; blue, 27 }  ,fill opacity=1 ] (178.3,141.07) .. controls (178.3,137.76) and (180.99,135.07) .. (184.3,135.07) .. controls (187.62,135.07) and (190.3,137.76) .. (190.3,141.07) .. controls (190.3,144.38) and (187.62,147.07) .. (184.3,147.07) .. controls (180.99,147.07) and (178.3,144.38) .. (178.3,141.07) -- cycle ;
\draw  [fill={rgb, 255:red, 208; green, 2; blue, 27 }  ,fill opacity=1 ] (209.3,141.07) .. controls (209.3,137.76) and (211.99,135.07) .. (215.3,135.07) .. controls (218.62,135.07) and (221.3,137.76) .. (221.3,141.07) .. controls (221.3,144.38) and (218.62,147.07) .. (215.3,147.07) .. controls (211.99,147.07) and (209.3,144.38) .. (209.3,141.07) -- cycle ;
\draw [line width=1.5]    (131.61,96.14) -- (240,96) ;
\draw  [fill={rgb, 255:red, 208; green, 2; blue, 27 }  ,fill opacity=1 ] (150,95.77) .. controls (150,92.46) and (152.69,89.77) .. (156,89.77) .. controls (159.31,89.77) and (162,92.46) .. (162,95.77) .. controls (162,99.09) and (159.31,101.77) .. (156,101.77) .. controls (152.69,101.77) and (150,99.09) .. (150,95.77) -- cycle ;
\draw  [fill={rgb, 255:red, 208; green, 2; blue, 27 }  ,fill opacity=1 ] (178.3,96.07) .. controls (178.3,92.76) and (180.99,90.07) .. (184.3,90.07) .. controls (187.62,90.07) and (190.3,92.76) .. (190.3,96.07) .. controls (190.3,99.38) and (187.62,102.07) .. (184.3,102.07) .. controls (180.99,102.07) and (178.3,99.38) .. (178.3,96.07) -- cycle ;
\draw  [fill={rgb, 255:red, 208; green, 2; blue, 27 }  ,fill opacity=1 ] (209.3,96.07) .. controls (209.3,92.76) and (211.99,90.07) .. (215.3,90.07) .. controls (218.62,90.07) and (221.3,92.76) .. (221.3,96.07) .. controls (221.3,99.38) and (218.62,102.07) .. (215.3,102.07) .. controls (211.99,102.07) and (209.3,99.38) .. (209.3,96.07) -- cycle ;
\draw [fill={rgb, 255:red, 74; green, 144; blue, 226 }  ,fill opacity=1 ][line width=0.75]    (352.09,97.02) -- (352.09,107.79) ;
\draw [fill={rgb, 255:red, 74; green, 144; blue, 226 }  ,fill opacity=1 ][line width=0.75]    (321.3,96.07) -- (321.3,107.84) ;
\draw [fill={rgb, 255:red, 74; green, 144; blue, 226 }  ,fill opacity=1 ][line width=0.75]    (293,95.77) -- (293,140.55) ;
\draw [line width=1.5]    (268.61,141.14) -- (377,141) ;
\draw  [fill={rgb, 255:red, 208; green, 2; blue, 27 }  ,fill opacity=1 ] (287,140.77) .. controls (287,137.46) and (289.69,134.77) .. (293,134.77) .. controls (296.31,134.77) and (299,137.46) .. (299,140.77) .. controls (299,144.09) and (296.31,146.77) .. (293,146.77) .. controls (289.69,146.77) and (287,144.09) .. (287,140.77) -- cycle ;
\draw [fill={rgb, 255:red, 74; green, 144; blue, 226 }  ,fill opacity=1 ][line width=0.75]    (321.3,130.18) -- (321.3,140.96) ;
\draw  [fill={rgb, 255:red, 208; green, 2; blue, 27 }  ,fill opacity=1 ] (315.3,141.07) .. controls (315.3,137.76) and (317.99,135.07) .. (321.3,135.07) .. controls (324.62,135.07) and (327.3,137.76) .. (327.3,141.07) .. controls (327.3,144.38) and (324.62,147.07) .. (321.3,147.07) .. controls (317.99,147.07) and (315.3,144.38) .. (315.3,141.07) -- cycle ;
\draw  [fill={rgb, 255:red, 208; green, 2; blue, 27 }  ,fill opacity=1 ] (346.3,141.07) .. controls (346.3,137.76) and (348.99,135.07) .. (352.3,135.07) .. controls (355.62,135.07) and (358.3,137.76) .. (358.3,141.07) .. controls (358.3,144.38) and (355.62,147.07) .. (352.3,147.07) .. controls (348.99,147.07) and (346.3,144.38) .. (346.3,141.07) -- cycle ;
\draw [line width=1.5]    (268.61,96.14) -- (377,96) ;
\draw  [fill={rgb, 255:red, 208; green, 2; blue, 27 }  ,fill opacity=1 ] (287,95.77) .. controls (287,92.46) and (289.69,89.77) .. (293,89.77) .. controls (296.31,89.77) and (299,92.46) .. (299,95.77) .. controls (299,99.09) and (296.31,101.77) .. (293,101.77) .. controls (289.69,101.77) and (287,99.09) .. (287,95.77) -- cycle ;
\draw  [fill={rgb, 255:red, 208; green, 2; blue, 27 }  ,fill opacity=1 ] (315.3,96.07) .. controls (315.3,92.76) and (317.99,90.07) .. (321.3,90.07) .. controls (324.62,90.07) and (327.3,92.76) .. (327.3,96.07) .. controls (327.3,99.38) and (324.62,102.07) .. (321.3,102.07) .. controls (317.99,102.07) and (315.3,99.38) .. (315.3,96.07) -- cycle ;
\draw  [fill={rgb, 255:red, 208; green, 2; blue, 27 }  ,fill opacity=1 ] (346.3,96.07) .. controls (346.3,92.76) and (348.99,90.07) .. (352.3,90.07) .. controls (355.62,90.07) and (358.3,92.76) .. (358.3,96.07) .. controls (358.3,99.38) and (355.62,102.07) .. (352.3,102.07) .. controls (348.99,102.07) and (346.3,99.38) .. (346.3,96.07) -- cycle ;

\draw (145,149.17) node [anchor=north west][inner sep=0.75pt]    {$A$};
\draw (175,149.4) node [anchor=north west][inner sep=0.75pt]    {$A$};
\draw (206,149.17) node [anchor=north west][inner sep=0.75pt]    {$A$};
\draw (147,73.4) node [anchor=north west][inner sep=0.75pt]    {$\bar{A}$};
\draw    (148,108) -- (190,108) -- (190,130) -- (148,130) -- cycle  ;
\draw (151,112.4) node [anchor=north west][inner sep=0.75pt]  [font=\small]  {$\ \ \ X\ \ \ $};
\draw (175,73.4) node [anchor=north west][inner sep=0.75pt]    {$\bar{A}$};
\draw (205,73.4) node [anchor=north west][inner sep=0.75pt]    {$\bar{A}$};
\draw    (317,108) -- (359,108) -- (359,130) -- (317,130) -- cycle  ;
\draw (320,112.4) node [anchor=north west][inner sep=0.75pt]  [font=\small]  {$\ \ \ X\ \ \ $};
\draw (247,115.4) node [anchor=north west][inner sep=0.75pt]    {$=$};
\draw (283,149.17) node [anchor=north west][inner sep=0.75pt]    {$A$};
\draw (313,149.4) node [anchor=north west][inner sep=0.75pt]    {$A$};
\draw (344,149.17) node [anchor=north west][inner sep=0.75pt]    {$A$};
\draw (286,73.4) node [anchor=north west][inner sep=0.75pt]    {$\bar{A}$};
\draw (314,73.4) node [anchor=north west][inner sep=0.75pt]    {$\bar{A}$};
\draw (344,73.4) node [anchor=north west][inner sep=0.75pt]    {$\bar{A}$};
\end{tikzpicture}
\end{equation}
where $\bar{A}$ denotes the complex conjugate of $A$. 
Following this, we define the generalized transfer matrix as an operator $O$ sandwiched by $r$-local MPS blocks:
\begin{equation}
    T^{(r)}_O\equiv \sum_{\substack{a_1,\dots,a_r\\ b_1,\dots,b_r} }(O_{a_1\dots a_r}^{b_1\dots b_r})A^{[a_1]}\cdots A^{[a_r]}\otimes \bar{A}^{[b_1]}\cdots \bar{A}^{[b_r]}.
\end{equation}
With this notation, we are able to rewrite Eq.~\eqref{eq:transfer_matrix_graph} as
\begin{equation}\label{eq:transfer_matrix_comm}
   T^{(r-1)}_X T_\mathbb{I}^{(1)} =  T_\mathbb{I}^{(1)}T^{(r-1)}_X \implies [T^{(r-1)}_X, T_\mathbb{I}^{(1)}] = 0.
\end{equation}
It follows that $T^{(r-1)}_X$, if not zero, must belong to the commutant of $T_\mathbb{I}^{(1)}$.
The solution space is thus the inverse image of the commutant under the map $X\mapsto T^{(r-1)}_X$, which also contains the kernel.
Next, we analyze redundancies in the solution space, for which there are three possible mechanisms.
The first one is the trivial solution $X \propto i \mathbb{I}$, for which Eq.~\eqref{eq:reconstruction} leads to zero.
Second, solutions in the kernel $T^{(r-1)}_X = 0$ are redundant.
In this case, the local Hermitian term constructed from Eq.~\eqref{eq:reconstruction} is itself a telescoping term up to the addition of local type I Hamiltonians, and therefore sums to zero over the lattice~\cite{Supp}.
Third, when the transfer matrix $T^{(1)}_\mathbb{I}$ contains a zero-eigenvalue sector, including possible Jordan blocks, its commutant may contain elements $T^{(r-1)}_X$ satisfying $T^{(r-1)}_X \left(T^{(1)}_\mathbb{I}\right)^m = 0$.
Such a solution $X$ can be extended by $m$ sites according to: $\tilde{X}_{1:r+m-1} = X_{1:r-1}\otimes \mathbb{I}_{r:r+m-1}$, for which $T^{(r+m-1)}_{\tilde{X}} = 0$, thereby reducing to the second mechanism.
We now fix an operator range $r\ge l_\text{inj} + 1$.
By injectivity, the image of the map $X\mapsto T^{(r-1)}_X$, with anti-Hermitian $X$, becomes stable once this range is reached (see~\cite{Supp}).
For every element $\mathcal{E}$ in the intersection of this image and the commutant of $T^{(1)}_\mathbb{I}$, one can find a solution $X$ such that $T_X^{(r-1)} = \mathcal{E}$.
Therefore, after excluding the redundancies described above, the number of independent type II Hamiltonians is independent of $r$.
At first sight, this range independence may appear surprising.
However, note that the solution spaces are naturally nested as $r$ increases: a solution $X$ with range $(r-1)$ can be promoted to a range-$r$ solution $X\otimes \mathbb{I}$.
Consequently, once the number of solutions stops growing with $r$, no genuinely new inequivalent type II Hamiltonians can appear at larger range.
Therefore, all inequivalent type II Hamiltonians are exhausted at $r = l_\text{inj} + 1$. 
Note that previous numerical approaches proposed for finding such generalized parent Hamiltonians incur a computational cost growing exponentially with the operator range $r$, and therefore require an artificial cutoff that is rather small in practice~\cite{Qi2019Determining, moudgalya2023numerical, Wang2024Generalized, Gioia2026Distinct}.
In contrast, our theory establishes the exact number.
Here we present a counting formula for the number of inequivalent type II Hamiltonians, which turns out to depend only on the spectrum of $T^{(1)}_\mathbb{I}$. 
For simplicity, we assume that it is diagonalizable; the non-diagonalizable case, together with the general derivation, is discussed in \cite{Supp}.
If the spectrum of the transfer matrix $T^{(1)}_\mathbb{I}$ consists of eigenvalues $\tau_\lambda$ each with the degeneracy $d_\lambda$, then the number of inequivalent type II Hamiltonians $\mathcal{N}$  reads
\begin{equation}\label{eq:counting}
    \mathcal{N} = (\sum_{\lambda:\tau_\lambda\neq 0 } d_\lambda^2)-1,
\end{equation}
where $\sum_{\lambda}d_\lambda = \chi^2$.
Here, this is simply the dimension of the commutant of $T^{(1)}_\mathbb{I}$, with the zero eigenvalues $\tau_\lambda=0$ not contributing to inequivalent solutions, as for reasons discussed above; the ``$-1$'' removes the trivial solution.
As an example, consider a random MPS whose transfer matrix has neither zero eigenvalues nor spectral degeneracies, one has $\mathcal{N} = \chi^2-1$.

\textit{Examples}---To gain further insight, we next study several physically relevant examples and show how the structure of the transfer matrix spectrum determines the number of inequivalent type II Hamiltonians.

\textit{Symmetric MPS:}
One important feature of MPS is that global symmetries can be encoded explicitly in the local structure.
We now show that symmetries can enforce degeneracies in the transfer matrix spectrum and increase the number of type II Hamiltonians.
Suppose that $\ket{\psi(A)}$ respects a global onsite symmetry group $G$: $U(g)^{\otimes L}\ket{\psi(A)}\propto\ket{\psi(A)}$, $g\in G$.
According to the fundamental theorem of MPS, the symmetry action is implemented locally as~\cite{Cirac2021Matrix}
\begin{equation}
    \sum_{b=1}^q U_{ab}(g) A^{[b]} = e^{i\phi(g)}v^\dagger(g)A^{[a]}v(g).
\end{equation}
That is, the local action of the linear representation $U(g)$ on the qudit is pushed through the tensor and converted into an action of the projective representation $v(g)$ on the virtual space. 
It follows that the transfer matrix satisfies
\begin{equation}
    [T_\mathbb{I}^{(1)}, v(g) \otimes \bar{v}(g)] = 0.
\end{equation}
The symmetry action $v(g) \otimes \bar{v}(g)$ is a linear representation of $G$ in the doubled virtual space.
Hence the transfer matrix is $G$-symmetric, and is constrained to take the form
\begin{equation}
    T^{(1)}_\mathbb{I} = \bigoplus_\rho \mathbbm{1}_{d_\rho} \otimes T_\rho,
\end{equation}
where $d_\rho$ are the dimensions of the irreducible representations of $G$.
Consequently, every eigenvalue of $T_\rho$ is at least $d_\rho$-fold degenerate in the full spectrum.
For a non-Abelian group $G$, some non-zero sectors necessarily have $d_\rho>1$, thereby enforcing spectral degeneracies and enlarging the commutant of $T_{\mathbb{I}}^{(1)}$.
According to our counting formula Eq.~\eqref{eq:counting}, it leads to more type II Hamiltonians compared with non-symmetric cases.
For a paradigmatic example, we consider the ground state of the spin-1 Affleck-Kennedy-Lieb-Tasaki (AKLT) model under PBC~\cite{Affleck1988Valence, Verstraete2004Diverging}, which has $q=3$ and $\chi=2$, and the local tensors $\{A^{[a]}\}$ are proportional to the three Pauli matrices.
The state respects the non-Abelian $\mathrm{SO}(3)$ spin rotation symmetry, which is represented projectively as spin-$1/2$ on the virtual space.
The corresponding doubled virtual space decomposes into one singlet and one triplet, with $d_0 = 1$ and $d_1 = 3$, respectively.
Following Eq.~(\ref{eq:counting}), we therefore obtain $\mathcal{N} = 9$ inequivalent type II Hamiltonians.
We explicitly construct them in \cite{Supp}, which we find to be the three global spin rotation generators $S^{x,y,z}$~\cite{Moudgalya2022Exhaustive, Wang2024Generalized, Gioia2026Distinct} and the scalar spin chirality identified previously~\cite{sanada2023quantum}, together with five newly identified Hamiltonians. The latter two do not work under open boundary conditions.
A representative example has the two-site local Hermitian term: \(h_{j,j+1} = (S_j^z+S_{j+1}^z)(S_j^x S_{j+1}^y - S_j^y S_{j+1}^x)\).

\textit{Zero correlation length states:}
Connected correlation functions of an injective MPS decay exponentially with distance.
A representative subclass is zero correlation length states, where correlations vanish exactly beyond a finite distance $m$~\cite{Cirac2017MPDO}.
This subclass includes product states and states obtained from them by finite-depth local quantum circuits or quantum cellular automata.
Such states are also referred to as fixed-point states, since after an appropriate normalization, their transfer matrix satisfies~\cite{Cirac2017MPDO}
\begin{equation}
    \left(T^{(1)}_\mathbb{I}\right)^{m+l} = \left(T^{(1)}_\mathbb{I}\right)^{m},\quad \forall l>0.
\end{equation}
In other words, after grouping sufficiently many sites, the transfer matrix reduces to a rank-one projector, and the spectrum contains a single nonzero eigenvalue.
According to the counting formula, we then have $\mathcal{N} = 0$ and therefore no type II Hamiltonians, consistent with the result obtained in Ref.~\cite{Gioia2026Distinct} using a different approach.

\textit{Block-injective MPS}---
We now go beyond the injective TI MPS and consider the more general case of block-injective TI MPS.
The local tensor takes the form $A^{[a]} = \bigoplus_{\alpha=1}^\Lambda A^{[a]}_\alpha$, where within each $\alpha$-block the set of $\chi_\alpha$-dimensional matrices $A^{[a]}_\alpha$ is injective~\cite{Perez2007Matrix}.
We require the blocks to be inequivalent, namely, they cannot be related by phase and similarity transformations. Correspondingly, the state can be rewritten as $\ket{\psi(A)} = \sum_{\alpha=1}^\Lambda \ket{\psi(A_\alpha)}$, where each component is an injective TI MPS.
Indeed, a general homogeneous TI MPS as defined in Eq.~\eqref{eq:TI MPS} can be brought into such a form~\cite{Perez2007Matrix, Cirac2017MPDO}. 
In contrast to an injective MPS, the superposition structure allows nonvanishing connected correlation functions at arbitrarily large distances.
This property makes block-injective MPS a natural framework for describing spontaneous symmetry breaking, for example the Greenberger-Horne-Zeilinger (GHZ) state~\cite{Raussendorf2001One}.
As for the generalized parent Hamiltonians, the absence of type III Hamiltonians is not contained in prior work~\cite {Rubio2026Local}.
To fill this gap, in \cite{Supp}, we develop a local characterization to show that the generalized parent Hamiltonian must have all injective components as degenerate eigenstates, and provide a rigorous proof for the absence of type III Hamiltonians for block-injective MPS.
The crucial distinction brought by the block structure is that the map $X\mapsto T^{(r-1)}_X$ at large $r$ is no longer surjective onto the full matrix algebra of the doubled virtual space, but only onto diagonal blocks in the doubled virtual space, a property known as \textit{block-injectivity}~\cite{Cirac2021Matrix}. 
After a detailed analysis~\cite{Supp}, we obtain a similar formula for the number of inequivalent type II Hamiltonians:
\begin{equation}
    \mathcal{N} = (\sum_{\alpha\beta} \sum_{\tau_{\lambda,\alpha\beta}\neq 0}d_{\lambda,\alpha\beta}^2)-2\Lambda+k,
\end{equation}
where $(\tau_{\lambda,\alpha\beta},d_{\lambda,\alpha\beta})$ denote eigenvalues and degeneracies in the spectrum of the $(\alpha\beta)$-block transfer matrix:
\begin{equation}
    T_{\mathbb{I},\alpha\beta}^{(1)} = \sum_{a=1}^q A_\alpha^{[a]}\otimes \bar{A}_\beta^{[a]},
\end{equation}
and $k$ is the largest number of sets the $r$-local MPS subspaces associated with the different blocks can be divided into, such that subspaces in different sets are mutually orthogonal~\cite{Supp}.
For a random block-injective MPS, we expect all components to be non-orthogonal to each other, hence $k=1$, which gives $\mathcal{N} = (\sum_{\alpha=1}^\Lambda\chi_\alpha)^2-2\Lambda+1$.
Consider the GHZ state as a simple example.
The block-injective representation has two blocks $A_1^\uparrow =A_2^\downarrow = 1$, $A_1^\downarrow =A_2^\uparrow = 0$. The transfer matrices in $(11)$ and $(22)$ blocks each contain one nonzero eigenvalue, whereas $(12)$ and $(21)$ are zero and give no contribution. 
The two injective components are orthogonal spin-polarized states, corresponding to $k=2$.
Subtracting $2\Lambda-k=2$ therefore gives $\mathcal{N}=0$, showing that the GHZ state admits no type II Hamiltonian.
We also consider the two degenerate ground states of the Majumdar-Ghosh model~\cite{Majumdar1969NNN}. Their superposition admits a block-injective MPS representation, for which we obtain a complete characterization of $\mathcal{N}=7$ inequivalent type II Hamiltonians in \cite{Supp}.

\textit{Conclusions and outlook}---
In summary, we have developed a systematic method for constructing generalized parent Hamiltonians of TI MPS.
Our approach goes beyond conventional parent Hamiltonian constructions by relaxing the frustration-free condition and allowing the action of the local Hamiltonian on the state to leave non-Hermitian telescoping terms that cancel out when summed over.
For the injective case, we derived a local tensor equation that provides a systematic characterization and counting of the equivalence classes of such generalized parent Hamiltonians, which are determined solely by the MPS transfer matrix.
We illustrated the method through several examples to demonstrate how different physical properties affect the structure of generalized parent Hamiltonians.
We further extended the approach to the more general class of block-injective MPS.
The full characterization of type II Hamiltonians opens several directions for future exploration.
First, it would be interesting to study whether type I and type II Hamiltonians here differ in some nonequilibrium dynamics for certain initial states related to the target MPS, as observed in certain non-MPS examples~\cite{Gioia2026Distinct}.
Moreover, although not constructed as such, a type II Hamiltonian can in principle have the target MPS as its ground state, and it would be interesting to understand if and how the properties of such a generalized parent Hamiltonian differ from type I frustration-free Hamiltonians that are usually studied in the literature.

Future work should also consider MPS with open boundary conditions, where uncanceled telescoping terms at the boundaries can give rise to additional structures in the characterization of generalized parent Hamiltonians.
It would also be interesting to extend the present framework from a single MPS to a set of excited states on top of one, which can be formally captured using the recent MPS with a boundary~\cite{Floridollinas2025Uniform}.
This is particularly relevant to several QMBS models, where the set of states exhibits a tower structure in the sense that the scar states are generated by repeated action of a raising operator $R$ on a vacuum state: $\ket{\psi_n} = R^n\ket{\psi_0}$~\cite{Moudgalya2022Quantum, Chandran2022Quantum}.
Our studies of block-injective MPS already point to the additional complexity in the case of multiple states, suggesting a substantially richer structure of the generalized parent Hamiltonian problem.

Beyond specific states, the techniques developed here also suggest a natural extension to open quantum systems where the construction of parent Lindbladians of Matrix Product Density Operators (MPDO) is an important question. 
Previous constructions have largely focused on frustration-free parent Lindbladians, in which each local dissipative term individually annihilates target MPDOs~\cite{Verstraete2009Quantum, Bondarenko2021Constructing, Wang2024Embedding, Liu2026Parent, Baruah2026Dissipative}.
Our framework might help construct ``type-II'' Lindbladians while simultaneously enforcing complete positivity and trace preservation, which may have interesting properties and provide fresh insights for dissipative state-engineering.

\textit{Acknowledgments}---
We thank Marta Florido Llin\`{a}s, Andrew Ivanov, Olexei Motrunich, and Georgios Styliaris  for helpful discussions.
H.~R.~W. thanks Dong Yuan for a related collaboration~\cite{Wang2024Generalized}, and thanks Zhong Wang for hospitality at Tsinghua University.
S.~M. thanks Lei Gioia and Olexei Motrunich for a related collaboration~\cite{Gioia2026Distinct}.
H.~R.~W. acknowledges support from the National Natural Science Foundation of China under Grant No. 12125405 and the National Key R\&D Program of China (No. 2023YFA1406702).
L.G. acknowledges funding from the Swiss National Science Foundation under grant number 200020-219400.
S.M. acknowledges support from the Munich Center for Quantum Science and Technology (MCQST), which is supported by the Deutsche Forschungsgemeinschaft (DFG, German Research Foundation) under Germany’s Excellence Strategy--EXC--2111--390814868.

\bibliographystyle{apsrev4-1-title}
\bibliography{Heran_scar}

\end{document}


\title{Supplemental Material for: Exhaustive Construction of Generalized Parent Hamiltonians for Matrix Product States}

\author{He-Ran Wang}
\email{heranwang7@gmail.com}
\affiliation{Chen-Ning Yang Institute for Advanced Study, Tsinghua University, Beijing 100084,
People's Republic of China}

\author{Lorenzo Gotta}
\email{lorenzo.gotta@unige.ch}
\affiliation{Department of Quantum Matter Physics, University of Geneva, 1211 Geneva, Switzerland}

\author{Sanjay Moudgalya}
\email{sanjay.moudgalya@gmail.com}
\affiliation{Department of Theoretical Physics, Tata Institute of Fundamental Research, Mumbai 400005, India}
\affiliation{Technical University of Munich, TUM School of Natural Sciences, Physics Department, 85748 Garching, Germany}
\affiliation{Munich Center for Quantum Science and Technology (MCQST), Schellingstr. 4, 80799 M{\"u}nchen, Germany}

\maketitle

\tableofcontents
\section{Equivalence classes of Hamiltonians and redundant solutions}\label{sec:equivalence_and_redundancy}

We first briefly review the concept of equivalence classes of parent Hamiltonian operators~\cite{Moudgalya2022Exhaustive, Gioia2026Distinct}.
Consider, for example, the set of type II Hamiltonians for a given state. 
At first sight, one might expect them to form a linear space, since linear combinations preserve the type label. 
However, there is an important subtlety since two type II Hamiltonians can differ only by a type I Hamiltonian, which breaks the closure property according to the exclusive definition of the three types.
It is thus natural to regard such two type II Hamiltonians as equivalent, and characterize the linear space of inequivalent type II Hamiltonians.
This is also a more interesting question since the number of type I Hamiltonians grows exponentially with range, and we clearly understand their structure; hence we are interested only in the new parent Hamiltonians modulo the type I ones.
The same consideration applies to type III Hamiltonians, for which the equivalence relation is defined as addition of type II Hamiltonians.
We now characterize the type I equivalence in terms of the corresponding non-Hermitian telescoping operators.
In the main text, we established that a type II Hamiltonian is associated with the non-Hermitian telescoping term $X$ satisfying the transfer-matrix equation:
\begin{equation}\label{S_eq:transfer_matrix_comm}
    [T_X^{(r-1)}, T_\mathbb{I}] = 0.
\end{equation}
Consider two equivalent type II Hamiltonians,
\begin{equation}
    H_\mathrm{II} = \sum_j h_{j:j+r-1},\qquad H'_\mathrm{II} = \sum_j h'_{j:j+r-1}.
\end{equation}
associated with non-Hermitian telescoping operators $X$ and $X'$, respectively.
In this section, we address the question of what additional condition is satisfied by their difference $\delta X=X-X'$, given that the two Hamiltonians are equivalent.
Note that by linearity, $\delta X$ is naturally a solution of Eq.~\eqref{S_eq:transfer_matrix_comm}.
We are therefore asking which solutions correspond to type I equivalence and should be regarded as redundant.
To this end, we write the type I Hamiltonian given by the difference of two equivalent type II Hamiltonians:
\begin{equation}
    H_I = H_\mathrm{II}-H'_\mathrm{II} = \sum_j \delta h_{j:j+\tilde{r}-1},
\end{equation}
where each term $\delta h_{j:j+\tilde{r}-1}$ annihilates the MPS locally.
In general, the range $\tilde r$ of the type I Hamiltonian need not coincide with $r$.
If $\tilde r<r$, the shorter-range terms can always be embedded into an $r$-site support, and we may therefore take $\tilde r\ge r$ without loss of generality.
Note that the relation between the local terms does not follow directly from that of global Hamiltonians.
As shown in Ref.~\cite{Rubio2026Local}, different local representations of the same global Hamiltonian may differ by additional telescoping terms.
We can therefore write
\begin{equation}\label{eq:difference_of_typeII}
    h_{j:j+r-1} -  h'_{j:j+r-1} = \delta h_{j:j+\tilde{r}-1} +g_{j:j+\tilde{r}-2} -g_{j+1:j+\tilde{r}-1},
\end{equation}
where $g$ is Hermitian.
Acting with this relation on the MPS and using Eq.~(2) of the main text, we obtain
\begin{equation}
    (\delta X_{j:j+r-2}+g_{j:j+\tilde{r}-2})\ket{\psi(A)} = (\delta X_{j+1:j+r-1}+g_{j+1:j+\tilde{r}-1})\ket{\psi(A)}.
\end{equation}
It follows that the operator 
\begin{equation}\label{eq:Y_operator}
   Y_{1:\tilde r-1} = \delta X_{1:r-1}\otimes I_{r:\tilde r-1} + g_{1:\tilde r-1}
\end{equation}
has the same action on the MPS when translated from one position to the next. Here $I$ is the identity operator.
$\delta X$ and $g$ are the anti-Hermitian and Hermitian part of $Y$, respectively. 
By repeatedly applying this relation, the operator $Y$ on two sides of the equation can be separated by an arbitrary distance.
We then discuss how this counting proceeds for the injective and block-injective MPS cases separately.
\subsection{Injective MPS}\label{sec:equivalence_and_redundancy_inj}
For an injective MPS, this leads to a strong constraint, as given by the following lemma:
\begin{lemma}\label{lemma:injective_separate}
Let $\ket{\psi(A)}$ be an injective MPS with system size $L$ and injectivity length $l_\mathrm{inj}$.
If an operator $Y$ of range $r$ satisfies 
\begin{equation}
    Y_{j:j+r-1}\ket{\psi(A)}=Y_{j+l:j+l+r-1}\ket{\psi(A)},
\end{equation}
with $l-r\ge l_\mathrm{inj}, L-l-r\ge l_\mathrm{inj}$, then $Y$ acts as a scalar on the subspace spanned by $r$-local MPS:
\begin{equation}
  \exists c\in\mathbb{C},\quad  Y_{1:r} \ket{A^{(r)}_{\mu\nu}} = c \ket{A^{(r)}_{\mu\nu}},\quad \mu, \nu = 1, 2, \dots, \chi.
\end{equation}
\end{lemma}
\begin{proof}
Under PBC, insertions of two $Y$ operators divide the remaining part of the MPS into two regions of lengths $l-r$ and $L-l-r$, respectively.
By assumption, both lengths are no smaller than $l_{\mathrm{inj}}$.
Injectivity therefore allows us to apply an inverse map to each of these two regions, mapping the physical degrees of freedom to the virtual space at the region boundaries.
Applying these inverse maps to the equality removes tensors in the two regions and yields
\begin{equation}
    \left(Y_{1:r} \ket{A^{(r)}_{\mu\nu}}\right)\otimes\ket{A^{(r)}_{\mu'\nu'}} = \ket{A^{(r)}_{\mu\nu}}\otimes \left(Y_{1:r} \ket{A^{(r)}_{\mu'\nu'}}\right).
\end{equation}
Choosing an arbitrary state $\ket{\varphi}$ and acting with $I\otimes\bra{\varphi}$ on the above equality, one can prove the lemma.
\end{proof}
This proves that the operator $Y_{1:\tilde r-1}$ defined in Eq.~\eqref{eq:Y_operator} acts as a scalar on the subspace spanned by ($\tilde r - 1$)-local MPS. 
To characterize this redundancy in terms of $\delta X$, we follow a derivation analogous to Eqs.~(3)-(5) in the main text and obtain
\begin{align}\label{eq:injective_redudant}
    &\bra{A_{\mu' \nu' }^{(\tilde r-1)}}(Y_{1:\tilde r-1} - Y^\dagger_{1:\tilde r-1})\ket{A_{\mu \nu }^{(\tilde r-1)}} = (c-c^*) \bra{A_{\mu' \nu' }^{(\tilde r-1)}}A_{\mu \nu }^{(\tilde r-1)}\rangle\nonumber\\
    \implies & \bra{A_{\mu' \nu' }^{(\tilde r-1)}}(\delta X_{1:r-1}\otimes I_{r:\tilde r-1})\ket{A_{\mu \nu }^{(\tilde r-1)}} = i\ \text{Im}(c)\bra{A_{\mu' \nu' }^{(\tilde r-1)}}A_{\mu \nu }^{(\tilde r-1)}\rangle \nonumber\\
    \implies & T_{\delta X}^{(r-1)} \left(T_\mathbb{I}^{(1)}\right)^{\tilde r - r} = i\ \text{Im}(c) \left(T_\mathbb{I}^{(1)}\right)^{\tilde r - 1}.
\end{align}

The number of inequivalent type II Hamiltonians is therefore given by the dimension of the solution to the transfer matrix Eq.~\eqref{S_eq:transfer_matrix_comm}, minus the dimension of the redundant solution space.
To count the latter, we write $\delta X$ as $\delta X = i\ \text{Im}(c)\mathbb{I} + \delta X'$, and count the number of solutions to
\begin{equation}
    T_{\delta X'}^{(r-1)} \left(T_\mathbb{I}^{(1)}\right)^{\tilde r - r} = 0
\label{eq:redundancy_eqn}
\end{equation}
When $\tilde r = r$, this condition reduces to $\delta X'$ being in the kernel of the map $X\mapsto T^{(r-1)}_X$.
For $\tilde r > r$, additional redundant solutions can arise from the zero-eigenvalue sectors and zero-eigenvalue Jordan blocks of the transfer matrix $T_{\mathbb I}^{(1)}$.
Constructing the local Hamiltonian from such a $\delta X$ using Eq.~(6) of the main text yields precisely the structure on the right-hand side of Eq.~\eqref{eq:difference_of_typeII}, namely, a Hermitian telescoping term up to the addition of a type I Hamiltonian.

We illustrate the complete counting of solutions to the transfer matrix equation in Sec.~\ref{sec:injective_sol}.
\subsection{Block-Injective MPS}\label{sec:equivalence_and_redundancy_binj}
To extend the characterization to the block-injective case, we first review the notion of block-injectivity length.
Consider an MPS $\ket{\psi(A)}$ with $A = \bigoplus_\alpha^\Lambda A_\alpha$.
The block-injectivity length $l_{\mathrm{binj}}$ is defined as the minimum number of consecutive sites such that the products
$\{\prod_{j=1}^{l_{\mathrm{binj}}} A^{[\sigma_j]}\}$ span the full block-diagonal matrix algebra $\bigoplus_{\alpha=1}^{\Lambda} M_{\chi_\alpha}(\mathbb C)$ in the virtual linear space~\cite{Fernndez2014Frustration, Cirac2017MPDO}. 
In particular, upon grouping at least $l_{\mathrm{binj}}$ consecutive sites, the local MPS subspaces associated with different injective components are linearly independent.
Now we establish the following lemma:
\begin{lemma}\label{lemma:binjective_separate}
Let $\ket{\psi(A)}$ be a block-injective MPS with system size $L$ and block-injectivity length $l_\mathrm{binj}$. Suppose an operator $Y$ of range $r$ satisfies 
\begin{equation}
    Y_{j:j+r-1}\ket{\psi(A)}=Y_{j+l:j+l+r-1}\ket{\psi(A)},
\end{equation}
with $l-r\ge l_\mathrm{binj}, L-l-r\ge l_\mathrm{binj}$.
Then, for each injective component $A_\alpha$, $Y$ acts as a scalar on the subspace spanned by the corresponding $r$-local MPS:
\begin{equation}
  \forall \alpha, \quad \exists c_\alpha\in\mathbb{C},\quad  Y_{1:r} \ket{(A_\alpha)^{(r)}_{\mu\nu}} = c_\alpha \ket{(A_\alpha)^{(r)}_{\mu\nu}}, \quad \mu,\nu = 1, 2, \dots, \chi_\alpha.
\end{equation}
\end{lemma}

\begin{proof}
The proof is analogous to that of Lemma~\ref{lemma:injective_separate}.
The crucial difference is that block-injectivity allows us to resolve the individual block sectors in the virtual space and apply the corresponding inverse map within each block~\cite{Cirac2017MPDO}.
It follows that for any block $\alpha$ there is
\begin{equation}
    \left(Y_{1:r} \ket{(A_\alpha)^{(r)}_{\mu\nu}}\right)\otimes\ket{(A_\alpha)^{(r)}_{\mu'\nu'}} = \ket{(A_\alpha)^{(r)}_{\mu\nu}}\otimes \left(Y_{1:r} \ket{(A_\alpha)^{(r)}_{\mu'\nu'}}\right),
\end{equation}
thus completing the proof.
\end{proof}
We defer the counting of redundant solutions for $\delta X$ in this case to Sec.~\ref{sec:block_inj_sol}. 
\section{Solutions of the local tensor equation: Injective MPS}\label{sec:injective_sol}
In this section, we provide a detailed characterization of the solution space of Eq.~\eqref{S_eq:transfer_matrix_comm} for an injective MPS.
As discussed in the main text, we choose $r\ge l_\mathrm{inj} + 1$.
To characterize the image of the map $X\to T^{(r-1)}_X$, we first consider an arbitrary operator $O$.
By injectivity of the MPS, it is easy to show that the map $O\mapsto T^{(r-1)}_O$ is surjective onto the full matrix algebra in the doubled virtual space.
Moreover, by definition, $T^{(r-1)}_{O^\dagger} = \Theta \left[T^{(r-1)}_O\right]$, where $\Theta$ is an anti-linear operator that exchanges the two virtual spaces and performs complex conjugation on matrices in the doubled virtual space.
Restricting to anti-Hermitian operators therefore gives
\begin{equation}
    T^{(r-1)}_{O - O^\dagger} = T^{(r-1)}_{O} - T^{(r-1)}_{O^\dagger} = (I-\Theta)\left[T^{(r-1)}_O\right],
\end{equation}
whose image is precisely the $-1$ sector of $\Theta$. 
Recall that the full number of solutions to type II Hamiltonians is given in terms of generalized transfer matrices $\{T^{(r-1)}_X\}$ that are in the commutant of transfer matrix [see Eq.~\eqref{S_eq:transfer_matrix_comm}], which then needs to be reduced accounting for redundancies discussed in the previous section.
Define the commutant as
\begin{equation}
    \mathcal{C} = \{\mathcal{E}|[T_\mathbb{I}^{(1)}, \mathcal{E}] = 0\}.
\end{equation}
The commutant is a complex linear space, which is closed under linear combinations with complex coefficients.
On the other hand, the space of anti-Hermitian operators $X$ as solutions of the transfer matrix equation is a real linear space, which is only closed under linear combinations with real coefficients.
Nevertheless, in the following we show that the real dimension of the latter space equals the complex dimension of the former after accounting for the redundancies.
Since $\Theta \left[T_\mathbb{I}^{(1)}\right] = T_\mathbb{I^\dagger}^{(1)} = T_\mathbb{I}^{(1)}$, we find that for any commutant element $\mathcal{E}$, there is also
\begin{equation}
    [T_\mathbb{I}^{(1)}, \Theta \left[\mathcal{E}\right]] = 0,
\end{equation}
namely, $\mathcal{C}$ is invariant under $\Theta$.
Since $\Theta^2 = I$, we may decompose $\mathcal{C}$ into $+1$ and $-1$ sectors under $\Theta$.
We then need to determine the dimension of the $\Theta = -1$ sector of the commutant algebra, which is the number of equivalent classes of type II Hamiltonians after accounting for the redundancies, and forms a real linear space.
Consider a basis \(\{\mathcal E\}\) of the \(\Theta=-1\) sector.
It is straightforward to see that \(\{i\mathcal E\}\) forms a basis of the \(\Theta=+1\) sector, since multiplication by \(i\) gives a bijection between the two sectors.
Consequently, when allowing complex linear combinations, \({\mathcal E}\) spans the entire commutant algebra.
Therefore, the real dimension of the \(\Theta=-1\) sector equals the complex dimension of the commutant.
It follows that the dimension of the solution space is the sum of the dimension of the kernel of the map, where $T_X = 0$, and the dimension of $\mathcal{C}$.
In Sec.~\ref{sec:equivalence_and_redundancy_inj}, we have seen that any operator that satisfies Eq.~(\ref{eq:redundancy_eqn}) leads to redundancies, which include operators $X$ that are in the kernel.
It therefore remains to determine the dimension of $\mathcal{C}$, and identify the remaining redundancies therein, which, according to Eq.~(\ref{eq:redundancy_eqn}), can arise from zero eigenvalues in the transfer matrix.

\subsection{Commutant dimension: Diagonalizable case}\label{sec:injective_sol_diag}
The commutant algebra of a matrix becomes transparent from its spectral decomposition.
We first consider the diagonalizable case:
\begin{equation}
    T_\mathbb{I}^{(1)}= S \left(\bigoplus_\lambda \tau_\lambda \mathbbm{1}_{d_\lambda}\right) S^{-1},
\end{equation}
where $\lambda$ labels different eigenvalues $\tau_\lambda$, $d_\lambda$ denotes their multiplicities, and $S$ is a similarity transformation.
An operator $\mathcal{E}$ commutes with $T_{\mathbb I}^{(1)}$ if and only if it preserves each eigenspace,
\begin{equation}
[\mathcal{E},T_{\mathbb I}^{(1)}]=0
\quad\Longleftrightarrow\quad
\mathcal{E}= S\left( \bigoplus_\lambda \mathcal{E}_\lambda\right) S^{-1},
\end{equation}
where $\mathcal{E}_\lambda$ is an arbitrary matrix of dimension $d_\lambda$.
Therefore, each eigenspace labeled by $\lambda$ contributes $d_\lambda^2$ to the counting of the dimension of the commutant algebra. 
Suppose now that zero is an eigenvalue, and denote the corresponding eigenspace by $\lambda=0$, $\tau_0=0$.
An element in the commutant algebra supported entirely in this eigenspace takes the form
\begin{equation}
\mathcal{E}
=
S\left(
\mathcal{E}_0\oplus \bigoplus_{\lambda\neq0}0
\right)S^{-1},
\end{equation}
and there is a $d_0^2$-dimensional vector space of them.
For a solution satisfying $T_X^{(r-1)}=\mathcal{E}$, we have
\begin{equation}
T_X^{(r-1)}T_{\mathbb I}^{(1)}
=
S\left(
\tau_0 \mathcal{E}_0\oplus\bigoplus_{\lambda\neq0}0 \mathbbm{1}_{d_\lambda}
\right)S^{-1}
=0.
\end{equation}
These $d_0^2$ solutions are precisely the redundant solutions associated with the zero-eigenvalue sector discussed in Sec.~\ref{sec:equivalence_and_redundancy_inj}, since they satisfy Eq.~(\ref{eq:redundancy_eqn}), and should therefore be excluded from the counting.
In addition, the solution $X$ being proportional to the identity is also redundant which reduces the number by one more.
Hence, the number of equivalence classes of type II Hamiltonians is
\begin{equation}\label{S_eq:counting_formula_inj}
\mathcal{N} = (\sum_{\lambda:\tau_\lambda\neq0}d_\lambda^2)-1,
\end{equation}
which proves the counting formula Eq.~(11) of the main text.
\subsection{Commutant dimension: Non-Diagonalizable case}
For a non-diagonalizable transfer matrix, the full Jordan structure must be taken into account to analyze the commutant algebra.
We write the Jordan normal form of the transfer matrix as
\begin{equation}
    T_\mathbb{I}^{(1)}= S \left(\bigoplus_\lambda\bigoplus_{p=1}^{k_\lambda} J_{d_{\lambda,p}} (\tau_\lambda) \right) S^{-1},
\end{equation}
where $p=1,\ldots,k_\lambda$ labels the Jordan blocks associated with the eigenvalue $\tau_\lambda$, and $J_{d}(\tau)$ denotes a Jordan block of dimension $d$ with this eigenvalue. 
An operator $\mathcal{E}$ commuting with $T_{\mathbb I}^{(1)}$ still cannot mix different eigenspaces labeled by $\lambda$:
\begin{equation}
    \mathcal{E}= S\left( \bigoplus_\lambda \mathcal{E}_\lambda\right) S^{-1},
\end{equation}
but within a fixed eigenspace, $\mathcal{E}_\lambda$ can couple different Jordan blocks.
To determine the form of $\mathcal{E}_\lambda$, we parameterize it as
\begin{equation}
    \mathcal{E}_\lambda = \begin{pmatrix}
\mathcal{E}_{11} & \mathcal{E}_{12} & \cdots & \mathcal{E}_{1k_\lambda} \\
\mathcal{E}_{21} & \mathcal{E}_{22} & \cdots & \mathcal{E}_{2k_\lambda} \\
\vdots & \vdots & \ddots & \vdots \\
\mathcal{E}_{k_\lambda1} & \mathcal{E}_{k_\lambda2} & \cdots & \mathcal{E}_{k_\lambda k_\lambda}
\end{pmatrix},
\end{equation}
where the submatrices $\mathcal{E}_{pq}$ are $d_{\lambda,p}\times d_{\lambda,q}$ matrices.
The commuting relation demands that, for each pair of $(p,q)$, there is
\begin{equation}
    J_{d_{\lambda,p}}(\tau_\lambda)\mathcal{E}_{p,q} = \mathcal{E}_{p,q}J_{d_{\lambda,q}}(\tau_\lambda).
\end{equation}
Expanding this condition in matrix elements yields
\begin{equation}
    (\mathcal{E}_{p,q})_{m+1, n} = (\mathcal{E}_{p,q})_{m, n-1}, 
\end{equation}
together with boundary conditions
\begin{equation}
    (\mathcal{E}_{p,q})_{d_{\lambda,p}, n} = (\mathcal{E}_{p,q})_{m, 1}=0. 
\end{equation}
These relations imply that the matrix elements are constant along diagonals with a fixed difference between the column and row indices.
Among total $\left(d_{\lambda,p} + d_{\lambda,q}-1\right)$ diagonals, the boundary conditions set $\left(\max{(d_{\lambda,p}, d_{\lambda,q})}-1\right)$ of them to zero. 
Consequently, $\mathcal{E}_{p,q}$ contains $\min(d_{\lambda,p},d_{\lambda,q})$ non-vanishing diagonals, contributing this number to the commutant dimension.
We next consider the zero-eigenvalue sector, $\tau_0 = 0$.
In this sector, each Jordan block is nilpotent: $\left(J_{d_{0 ,p}}(\tau_0)\right)^{d_{0 ,p}} = 0$.
Consequently, for any element $\mathcal{E}$ in the commutant supported entirely in the zero-eigenvalue sector, we can choose $m=\max_p \{d_{0,p}\}$ such that
$\mathcal{E}\left(T_\mathbb{I}^{(1)}\right)^{m} = 0$.
According to Sec.~\ref{sec:equivalence_and_redundancy_inj},  in particular Eq.~(\ref{eq:redundancy_eqn}), all such solutions therefore correspond to redundant solutions and should be excluded from the counting.

In all, this leads us to the counting formula for the generic non-diagonalizable case:
\begin{equation}\label{S_eq:counting_formula_inj_Jordan}
    \mathcal{N} = \left(\sum_{\lambda:\tau_\lambda\neq0}\sum_{p,q}^{k_\lambda}\min(d_{\lambda,p},d_{\lambda,q})\right)-1.
\end{equation}
Eq.~\eqref{S_eq:counting_formula_inj} is recovered in the diagonalizable case, where all Jordan blocks have size one, i.e., $d_{\lambda, p} = 1$.

\section{Type II Hamiltonians for the AKLT ground state}\label{sec:AKLT}
Here, we present the $\mathcal{N}=9$ inequivalent type II Hamiltonians for the spin-$1$ AKLT ground state under PBC, which we know should exist by direct application of Eq.~(\ref{S_eq:counting_formula_inj}).
We work in the Cartesian basis $\ket{a}, a = x,y,z$ in which the local MPS tensors are proportional to the conventional Pauli matrices,
\begin{equation}
        A^{[a]}= -\sqrt{\frac{1}{3}} \sigma^a,\quad a=x,y,z,
\end{equation}
with injectivity length $l_\mathrm{inj} = 2$.
In this basis, the physical state $\ket{a}$ is the zero-eigenvalue state of one-site spin-$1$ operator $S^a$.
We seek solutions to $T_X^{(2)} =\mathcal{E}$, where $\mathcal{E}$ runs over elements in the commutant of $T_\mathbb{I}^{(1)}$. 
We begin by developing a systematic procedure for solving $T^{(2)}_X = \mathcal{E}$ in this case.
To eliminate redundant solutions, we restrict $X$ entirely to the support of $2$-local MPS, i.e., $Q^{(2)}XQ^{(2)} = X$, since for components outside this support, the generalized transfer matrix vanishes and is redundant as discussed in Sec.~\ref{sec:equivalence_and_redundancy_inj}.
It is convenient to construct a Cartesian basis for this four-dimensional subspace directly from $\ket{a}$ introduced above. We define
\begin{align}
&\ket{0}_{j,j+1}
=
\frac{1}{\sqrt{3}}
(
\ket{x}_j\ket{x}_{j+1}
+\ket{y}_j\ket{y}_{j+1}
+\ket{z}_j\ket{z}_{j+1}
),
&\ket{z}_{j,j+1}
=
\frac{i}{\sqrt{2}}
(
\ket{x}_j\ket{y}_{j+1}
-\ket{y}_j\ket{x}_{j+1})\nonumber\\
&\ket{x}_{j,j+1}=
\frac{i}{\sqrt{2}}
\left(
\ket{y}_j\ket{z}_{j+1}
-\ket{z}_j\ket{y}_{j+1}
\right),
&\ket{y}_{j,j+1}
=
\frac{i}{\sqrt{2}}
\left(
\ket{z}_j\ket{x}_{j+1}
-\ket{x}_j\ket{z}_{j+1}
\right).
\end{align}
Here, the total spin $J$ carried by $\ket{0}_{j,j+1}$ is zero, while $\ket{a}_{j,j+1}, a=x,y,z$ belong to the $J=1$ sector and are the zero-eigenvalue states of the corresponding total-spin components, similar to the one-site basis.
The \(J=2\) states lie outside the support subspace; indeed, the local terms of the AKLT Hamiltonian are projectors onto this sector.
We will represent $X$ in this two-site basis. The advantage of this representation becomes apparent from the form of the generalized transfer matrix, which can be simplified to be written as:
\begin{equation}
    T^{(2)}_X = \sum_{\substack{a_1a_2\\b_1b_2} }(X_{a_1a_2}^{b_1b_2})A^{[a_1]}A^{[a_2]}\otimes \bar{A}^{[b_1]} \bar{A}^{[b_2]} = \sum_{a, b = 0,x,y,z} X_a^b B^{[a]}\otimes \bar{B}^{[b]},
\end{equation}
where all the $B$ matrices are proportional to Pauli matrices:
\begin{equation}
B^{[0]} = \frac{1}{\sqrt{3}}\mathbb{I},\qquad B^{[x,y,z]} = \frac{\sqrt{2}}{3}\sigma^{x,y,z}.
\end{equation}
Consequently, the matrices $B^{[a]}\otimes \bar{B}^{[b]}$ form a complete orthogonal basis for matrices in the doubled virtual space.
This allows the coefficients of \(X\) to be extracted directly by projection. In particular, for $T^{(2)}_X = \mathcal{E}$, we obtain
\begin{equation}
    (X)_a^b = \frac{\mathrm{Tr}[(\bar{B}^{[a]}\otimes B^{[b]})\mathcal{E}]}{\mathrm{Tr}[B^{[a]}\bar{B}^{[a]}]\mathrm{Tr}[B^{[b]}\bar{B}^{[b]}]}.
\end{equation}
We next determine the commutant $\mathcal{C}$ of the transfer matrix $T_\mathbb{I}^{(1)}$, which reads
\begin{equation}
    T_\mathbb{I}^{(1)} = \frac{1}{3}[\sigma^x\otimes\sigma^x- \sigma^y\otimes\sigma^y + \sigma^z\otimes\sigma^z].
\end{equation}
For convenience, we introduce the normalized vectorized operators as states in the doubled virtual space:
\begin{equation}\label{eq:vectorized_operator}
    |O )\!)\equiv \frac{\sum_{\mu\nu}O_{\mu\nu}|\mu)\otimes|\nu)}{\sqrt{\sum_{\mu\nu} |O_{\mu\nu}|^2}}.
\end{equation}
In this basis, the spectral decomposition of the transfer matrix takes the form
\begin{equation}
    T_\mathbb{I}^{(1)} = |\mathbb{I})\!)(\!(\mathbb{I}| - \frac{1}{3}\left(|\sigma^x)\!)(\!(\sigma^x|+|\sigma^y)\!)(\!(\sigma^y|+|\sigma^z)\!)(\!(\sigma^z|\right).
\end{equation}
Thus, the transfer matrix contains a one-dimensional singlet sector with eigenvalue $1$ and a threefold-degenerate sector with eigenvalue $-1/3$, consistent with the symmetry analysis.
According to Sec.~\ref{sec:injective_sol_diag}, its commutant consists of the projector onto $|\mathbb{I})\!)$, together with the full matrix algebra in the subspace spanned by $|\sigma^x)\!), |\sigma^y)\!)$ and $|\sigma^z)\!)$.
This is hence $10$-dimensional, which ultimately leads to $9$ equivalence classes of type II Hamiltonians, from Eq.~(\ref{S_eq:counting_formula_inj}).
Finally, we describe how we convert the resulting \(X\) into local Hermitian terms of type II Hamiltonians.
In the main text, Eq.~(6) provides a general construction for this purpose.
Although formally applicable to arbitrary MPS, directly applying this construction to the AKLT state does not yield the most transparent expressions in terms of physical spin-\(1\) operators.
In particular, equivalent representatives of type II Hamiltonians may differ by type I Hamiltonians or by Hermitian telescoping terms, and an appropriate choice of these contributions can substantially simplify the expressions.
For the sake of elegance, we therefore propose a three-site Hermitian operator $h_{j,j+1,j+2}$ for each solution \(X_{j,j+1}\) and explicitly verify the corresponding local annihilation condition:
\begin{equation}
    \left(h_{1,2,3}+X_{1,2}-X_{2,3}\right)\ket{A_{\mu \nu }^{( 3)}} = 0.
\label{eq:typeIIfromgrad}
\end{equation}
Having introduced the solving technique and identified the commutant, we now present the inequivalent type II Hamiltonians obtained from the transfer matrix equation.
We organize the basis elements of the commutant according to how they transform under $\mathrm{SU}(2)$ symmetry, which naturally separates them into three categories: one-dimensional scalar, three-dimensional vector, and five-dimensional tensor.
As will be shown below, the corresponding type II Hamiltonians inherit the same transformation properties.
For each category, we present the $X$ solutions, the local Hermitian terms, and explicitly verify that they satisfy the required relation.
\subsection{Scalar spin chirality}~\label{sec:AKLT_scalar}
Scalars in the commutant consist of the projectors onto the singlet and triplet sectors, namely
\begin{equation}
    \mathcal{E}_0 = i |\mathbb{I})\!)(\!(\mathbb{I}|,\quad \mathcal{E}_1 = i (|\sigma^x)\!)(\!(\sigma^x|+ |\sigma^y)\!)(\!(\sigma^y|+|\sigma^z)\!)(\!(\sigma^z|).
\end{equation}
The two operators are invariant under $\mathrm{SU}(2)$ rotations on the doubled virtual space.
Both of them lead to $X$ operators that are linear combinations of two-site total spin-$0$ and total spin-$1$ projectors.
Since the sum of these two projectors acts as the identity on the two-local MPS subspace and corresponds to a type-I solution that should be removed, we choose one representative as the projector onto spin-$0$ on two sites, i.e.,
$X_{j,j+1} = i\ket{0}_{j,j+1} \bra{0}_{j,j+1}$ on sites $j$ and $j+1$.
Up to an overall factor, the corresponding type II Hamiltonian is the scalar spin chirality~\cite{Wen1989Chiral}:
\begin{equation}
    H = \sum\nolimits_j C_{j,j+1,j+2} = \sum\nolimits_j \vec{S}_j\cdot(\vec{S}_{j+1}\times \vec{S}_{j+2}),
\end{equation}
which is invariant under the global $\mathrm{SO}(3)$ rotations and therefore transforms as a scalar, like the commutant element $\mathcal{E}_0$.
This Hamiltonian has previously been identified as a generalized parent Hamiltonian of the AKLT ground state~\cite{Sanada2023Quantum}, while the type II-ness has not been shown previously.

Here, we establish its correspondence with the scalar anti-Hermitian solution by explicitly comparing their actions on the three-local MPS.
The key identity is
\begin{equation}\label{eq:spin_in_tensor}
    S^{a}\ket{A} = \frac{1}{2}\ket{[A,\sigma^a]}\implies \bra{b}S^{a}\ket{A} = -i\sum_{c=x,y,z}\varepsilon_{abc}A^{[c]}.
\end{equation}
Here, $\ket{A}$ is shorthand for \(\ket{A^{(1)}_{\mu\nu}}\), as defined in the main text, with the virtual indices suppressed. Similarly, we use \(\ket{A^{(r)}}\) to denote \(\ket{A^{(r)}_{\mu\nu}}\) for \(r>1\) throughout the following discussion.
Contracting with the physical operator and basis state leaves a matrix acting on the virtual indices.

The action of the anti-Hermitian telescoping term is
\begin{equation}
    \bra{abc}(X_{1,2} - X_{2,3})\ket{A^{(3)}} = \bra{abc}\left(i\ket{0}_{1,2}\ket{B^{[0]}A}-i\ket{AB^{[0]}}\ket{0}_{2,3}\right)=\frac{i}{3}\left(\delta_{ab}A^{[c]} - \delta_{bc}A^{[a]}\right).
\end{equation}
On the other hand, using Eq.~(\ref{eq:spin_in_tensor}) and the algebraic relations of the Pauli matrices, to which \(A^{[a]}\) are proportional, we can simplify the action of the scalar spin chirality as
\begin{align}
\bra{abc}C_{1,2,3}\ket{A^{(3)}} = &\sum_{def}\bra{abc}\varepsilon_{def} S_1^dS_2^eS_3^f\ket{A^{(3)}}\nonumber\\
=&\sum_{def}\sum_{ghi}  i \varepsilon_{def}\varepsilon_{dag}\varepsilon_{ebh}\varepsilon_{fci}A^{[g]}A^{[h]}A^{[i]}\nonumber\\
=&\frac{2i}{3}\left(\delta_{ab}A^{[c]} - \delta_{bc}A^{[a]}\right).
\end{align}
All the summations run over $x,y,z$. Therefore, we find that
\begin{equation}
    (-\frac{1}{2} C_{1,2,3} + X_{1,2} - X_{2,3})\ket{A^{(3)}} = 0,
\end{equation}
which completes the proof.

\subsection{Total spin operators}
Vectors in the commutant are given by the antisymmetric matrices acting within the triplet sector:
\begin{equation}
    \mathcal{E}_a
    =
    i\sum_{b,c = x,y,z}\varepsilon_{abc}
    |\sigma^b)\!)(\!(\sigma^c|,
    \qquad a=x,y,z.
\end{equation}
The three elements lead to two-site anti-Hermitian \(X\) operators parameterized by a real-valued vector $\vec{v}$:
\begin{equation}
    X_{j,j+1}
    =   
    \ket{0}_{j,j+1}\bra{\vec{v}}_{j,j+1}
    -
    \ket{\vec{v}}_{j,j+1}\bra{0}_{j,j+1},
    \qquad \ket{\vec{v}}_{j,j+1} = v_x\ket{x}_{j,j+1} + v_y\ket{y}_{j,j+1}+v_z\ket{z}_{j,j+1}.
\end{equation}
The corresponding type II Hamiltonians are the components of the total spin up to an overall factor:
\begin{equation}
    H=\sum\nolimits_j \vec{v}\cdot\vec{S}_j,
\end{equation}
which also transform as a vector under global $\mathrm{SO}(3)$ rotations.
Such Hamiltonians have already been identified as type II Hamiltonians of the AKLT ground state~\cite{Moudgalya2022Exhaustive, Wang2024Generalized, Gioia2026Distinct}.
To establish the correspondence, it is useful to introduce the two-site operator
\begin{equation}
    \widetilde X_{j,j+1}
    =
    6\ket{0}_{j,j+1}\bra{\vec{v}}_{j,j+1}
    +
    4\ket{\vec{v}}_{j,j+1}\bra{0}_{j,j+1},
\end{equation}
whose anti-Hermitian part is precisely \(X\).
Therefore, replacing \(X\) by \(\widetilde X\) amounts to adding a Hermitian telescoping term and leaves the global Hamiltonian unchanged.
Using the algebraic relations of the Pauli matrices, the action of the non-Hermitian telescoping term can be verified to be
\begin{align}
    \bra{abc}
    \left(
    \widetilde X_{1,2}
    -
    \widetilde X_{2,3}
    \right)
    \ket{A^{(3)}}
    = \frac{2\sqrt{6}}{3}\sum_d\left(-\sqrt{3}\delta_{ab} v_d A^{[d]}A^{[c]} + \sqrt{3}\delta_{bc} v_d A^{[a]}A^{[d]}
    + i \varepsilon_{abd}v_d A^{[c]} - i\varepsilon_{bcd}v_d A^{[a]}
    \right).
\end{align}
As for the three-site Hamiltonian, we choose the local term to be $h_{j, j+1, j+2} = \vec{v}\cdot(\vec{S}_j + \vec{S}_{j+2})$, which acts as
\begin{align}   
    \bra{abc}
    \left(\vec{v}\cdot(
    \vec{S}_1+\vec{S}_3)
    \right)
    \ket{A^{(3)}}
    =&\sum_{de} -i v_d \left(\varepsilon_{dae} A^{[e]}A^{[b]}A^{[c]} 
    + \varepsilon_{dce} A^{[a]}A^{[b]}A^{[e]}\right)\nonumber\\
    =& -\frac{1}{\sqrt{3}}\sum_d v_d \left(-\delta_{ab}A^{[d]}A^{[c]}+\delta_{bc}A^{[a]}A^{[d]}\right) -\frac{1}{3}\sum_d iv_d \left(\varepsilon_{abd}A^{[c]}-\varepsilon_{bcd}A^{[a]}\right).
\end{align}
We therefore obtain
\begin{equation}
    \left(2\sqrt{6} \vec{v}\cdot (\vec{S}_1+\vec{S}_3) + \widetilde X_{1,2} - \widetilde X_{2,3}\right)\ket{A^{(3)}} = 0,
\end{equation}
which completes the proof.
\subsection{Five new type II Hamiltonians}
The remaining five elements in the commutant transform as a symmetric traceless tensor under $\mathrm{SU}(2)$ spin rotations.
We parameterize them as
\begin{equation}
    \mathcal{E}=i\sum_{a,b=x,y,z}
    M_{ab}|\sigma^a)\!)(\!(\sigma^b|,
    \label{eq:aklt_tensor_commutant}
\end{equation}
where $M$ is a real symmetric traceless matrix.
It leads to the two-site $X$ operator
\begin{equation}
    X_{j,j+1}=i\sum_{a,b=x,y,z}M_{ab}\ket{a}_{j,j+1}\bra{b}_{j,j+1}.
\end{equation}
For later convenience, we replace $X$ by a spin-$1$ bilinear form
\begin{equation}
    \widetilde X_{j,j+1} = i \sum_{a,b=x,y,z}M_{ab}(S^a_j S^b_j +  S^a_{j+1} S^b_{j+1}).
\end{equation}
Using the properties of $M$, it can be verified that $T^{(2)}_X = T^{(2)}_{\widetilde{X}}$, namely, they correspond to the same equivalence class of type II Hamiltonians.

The corresponding type II Hamiltonians are given by
\begin{equation}
    H =\sum\nolimits_j g_{j,j+1} = \sum\nolimits_j i[\vec{S}_j\cdot\vec{S}_{j+1},\sum_{ab}M_{ab} \left(S_j^a S_j^b - S_{j+1}^a S_{j+1}^b\right)].
\end{equation}
The Hamiltonian $H$ transforms as a rank-two tensor under global $\mathrm{SO}(3)$ spin rotations. To our knowledge, this class of type II parent Hamiltonians for the AKLT ground state has not been identified previously.
As a representative component, consider
\begin{equation}
    M=\operatorname{diag}\left(-\frac{1}{3}, -\frac{1}{3}, \frac{2}{3}\right).
\end{equation}
The corresponding local Hermitian term is
\begin{equation}
    g_{j,j+1} = i[\vec{S}_j\cdot \vec{S}_{j+1}, (S_j^z)^2 - (S_{j+1}^z)^2] =\{S_j^z+S_{j+1}^z, S_j^x S_{j+1}^y - S_j^y S_{j+1}^x\} =2 (S_j^z+S_{j+1}^z)(S_j^x S_{j+1}^y - S_j^y S_{j+1}^x),
\end{equation}
which is the example presented in the main text.
Next, we prove the correspondence between $\widetilde X$ and $H$.
To this end, we make use of Eq.~\eqref{eq:spin_in_tensor} to establish the action of the bilinear form on the local tensor
\begin{align}
    \bra{c}\sum_{ab}M_{ab} S^a S^b \ket{A} 
    = -\sum_b M_{cb}A^{[b]}.
\end{align}
We have applied the traceless property of $M$ in the last equality.
It follows that the anti-Hermitian telescoping term acts on $\ket{A^{(3)}}$ as
\begin{align}\label{eq:X_on_A3}
    \bra{abc}\widetilde{X}_{1,2} - \widetilde{X}_{2,3}\ket{A^{(3)}} = -i\sum_d \left(M_{ad} A^{[d]} A^{[b]}A^{[c]}-M_{cd} A^{[a]}A^{[b]} A^{[d]}\right).
\end{align}
On the other hand, the action of the local Hermitian term $g_{1,2}$ on $\ket{A^{(2)}}$ is
\begin{equation}
    \bra{ab} g_{1,2}\ket{A^{(2)}}=-2i\sum_{e} \left(M_{ae} A^{[e]}A^{[b]} - M_{be} A^{[a]}A^{[e]}\right).
\end{equation}
Comparing it with Eq.~\eqref{eq:X_on_A3}, we find
\begin{equation}
    \left(-\frac{1}{2}g_{1,2}-\frac{1}{2}g_{2,3} + \widetilde{X}_{1,2} - \widetilde{X}_{2,3}\right)\ket{A^{(3)}}=0,
\end{equation}
which completes the proof.
\section{Absence of type III Hamiltonians for block-injective MPS}
In this section, we demonstrate the absence of type III Hamiltonians for block-injective MPS, and extend this result to a finite set of injective MPS.
\subsection{Local characterization of generalized parent Hamiltonians}
We first develop a local characterization of generalized parent Hamiltonians on block-injective MPS, generalizing the results of Ref.~\cite{Rubio2026Local} for injective MPS.
Intuitively, if a block-injective MPS is an eigenstate, one expects the Hamiltonian to act as a generalized parent Hamiltonian on each of its injective components.
To motivate this statement, consider the GHZ state as a simple example.
The GHZ state is a superposition of two mutually orthogonal spin-polarized product states, and it has been shown that any local generalized parent Hamiltonian of the GHZ state must have the two spin-polarized states as degenerate eigenstates~\cite{Facchi2011GHZ}.
Intuitively, the two spin-polarized states are locally orthogonal, so a local Hamiltonian cannot generate transitions between them, and therefore each component must itself be an eigenstate of the Hamiltonian. A general block-injective MPS exhibits an analogous structure, in the sense that upon grouping at least $l_{\mathrm{binj}}$ consecutive sites, the local MPS subspaces associated with different injective components become linearly independent.

We now provide the following theorem supporting the above observation:
\begin{theorem}\label{theorem}
Let $\ket{\psi(A)}$ be a block-injective MPS of system size $L$ and block-injectivity length $l_\mathrm{binj}$.
Any extensively local generalized parent Hamiltonian $H$ satisfying $H\ket{\psi(A)} = E\ket{\psi(A)}$, with local term of range at most $r$ and sufficiently large $L$, necessarily has all the injective components as degenerate eigenstates:
\begin{equation}
    H\ket{\psi(A_\alpha)} = E\ket{\psi(A_\alpha)},\quad\forall \alpha.
\end{equation}
\end{theorem}
\begin{proof}
Without loss of generality, we set $E=0$.
The proof proceeds by repeatedly using block-injectivity. 
For any $R\ge l_{\mathrm{binj}}$, block-injectivity implies that the $R$-local MPS subspaces associated with each block are linearly independent.
This allows one to construct a set of $R$-local operators $\Pi^{(R)}_\alpha$ for each subspace, satisfying
\begin{equation}
    \Pi^{(R)}_{\alpha, [j]}\ket{\psi(A)} = \ket{\psi(A_\alpha)}, \quad \Pi^{(R)}_{\alpha, [j]}\Pi^{(R)}_{\beta, [j]} = \delta_{\alpha\beta}\Pi^{(R)}_{\alpha, [j]}.
\end{equation}
These operators locally distinguish different sectors and are generally non-Hermitian, different from local projectors $Q^{(r)}$ defined in the main text.
Applying the projector $\Pi^{(R)}_\alpha$ to the eigenstate equation gives
\begin{equation}
    H \ket{\psi(A_\alpha)} = [H, \Pi^{(R)}_{\alpha, [j]}]\ket{\psi(A)}.
\end{equation}
The left-hand side is the action of $H$ on the injective component of interest. Since the local terms of $H$ are at most $r$-local, the commutator on the right-hand side is itself a local operator whose support has range at most $\tilde R\equiv2r+R-2$, denoted by $O_\alpha$.
Because the left-hand side is independent of the site $j$, the state generated by acting $O_\alpha$ on $\ket{\psi(A)}$ is shift-invariant, i.e., should be the same irrespective of the position of $O_\alpha$.
Lemma~\ref{lemma:binjective_separate} then implies that $O_\alpha$ acts as a scalar in each injective block:
\begin{equation}
    \forall \beta, \quad \exists c_{\alpha,\beta},\quad  O_{\alpha,1:\tilde R}  \ket{(A_\beta)^{(\tilde R)}_{\mu\nu}} = c_{\alpha,\beta} \ket{(A_\beta)^{(\tilde R)}_{\mu\nu}}, \quad \mu,\nu = 1, 2, \dots, \chi_\beta.
\end{equation}
At the level of the global MPS, we then have
\begin{equation}
   H \ket{\psi(A_\alpha)} = \sum_\beta^\Lambda c_{\alpha,\beta} \ket{\psi(A_\beta)}.
\end{equation}
To show that $\ket{\psi(A_\alpha)}$ is an eigenstate of $H$, it remains to show that all the coefficients vanish.
Take $\beta\neq\alpha$ and apply $\Pi^{(R)}_\beta$ to this equation:
\begin{equation}
     [\Pi^{(R)}_{\beta, [j]},H] \ket{\psi(A_\alpha)} = -O_{\beta,j:j+\tilde R-1}\ket{\psi(A_\alpha)} =  c_{\alpha,\beta}\ket{\psi(A_\beta)}.
\end{equation}
The right-hand side is again shift-invariant.
According to Lemma~\ref{lemma:injective_separate}, the local action of $O_\beta$ on $\ket{\psi(A_\alpha)}$ must be proportional to the same injective MPS:
\begin{equation}
    -O_{\beta,j:j+\tilde R-1}\ket{\psi(A_\alpha)} = c\ket{\psi(A_\alpha)} = c_{\alpha,\beta}\ket{\psi(A_\beta)}.
\end{equation}
Because $\ket{\psi(A_\alpha)}$ and $\ket{\psi(A_\beta)}$ are not the same state, we must have $c_{\alpha,\beta} = 0,\forall \beta\neq\alpha$.
Therefore, $\ket{\psi(A_\alpha)}$ becomes an eigenstate of $H$:
\begin{equation}
    H\ket{\psi(A_\alpha)} = c_{\alpha,\alpha}\ket{\psi(A_\alpha)}.
\end{equation}
Finally, combining the result with the eigenstate condition $H\ket{\psi(A)} = E\ket{\psi(A)}$ with $E = 0$, we have
\begin{equation}
    \sum_{\alpha=1}^\Lambda c_{\alpha,\alpha}\ket{\psi(A_\alpha)} = 0.
\end{equation}
Applying the projector $\Pi_\alpha^{(R)}$ on the equation proves that $c_{\alpha,\alpha} = 0$ for every $\alpha$.
\end{proof}
Theorem~\ref{theorem} extends the established ground-state degeneracy of a frustration-free parent Hamiltonian~\cite{Perez2007Matrix, Fernndez2014Frustration} for block-injective MPS to also generalized parent Hamiltonians of those MPS.
Since the generalized parent Hamiltonian $H$ has each injective component as an eigenstate, the individual action must be of type I or type II~\cite{Rubio2026Local}.
Namely, for each block $\alpha$, the local action can be written as
\begin{equation}
    h_{j:j+r-1}\ket{\psi(A_\alpha)} = (-Y_{\alpha,j:j+r-2}+Y_{\alpha,j+1:j+r-1})\ket{\psi(A_\alpha)}.
\end{equation}
The non-Hermitian telescoping operator $Y_\alpha$ may therefore depend on the block label $\alpha$. 
In contrast to the operator $X$ introduced in the main text, we do not assume a priori that the different $Y_\alpha$ can be chosen to be simultaneously anti-Hermitian.
This might suggest that unlike in the case of an injective MPS, block-injective MPS might admit a genuinely type III parent Hamiltonian when block-dependent operators $Y_\alpha$ cannot be chosen to be the same.
However, in the following, we show that this is not the case, and that $Y_\alpha$ can always be chosen to be block-independent.
\subsection{Existence of common non-Hermitian telescoping terms}
Fix an operator range $r\ge l_\mathrm{binj} + 1$, which can always be achieved by grouping local Hamiltonian terms.
As discussed before, block-injectivity allows us to construct a local operators  $\Pi_\alpha^{(r-1)}$ for each block, which annihilates injective components $\ket{\psi(A_\beta)}$ for $\beta\neq\alpha$.
Given a candidate type III Hamiltonian, we can combine the block-dependent telescoping terms $Y_\beta$ into a common operator:
\begin{equation}\label{eq:block_telescoping}
    X = \sum_{\alpha=1}^\Lambda Y_\alpha \Pi_\alpha^{(r-1)}.
\end{equation}
The defining property of $\Pi_\alpha^{(r-1)}$ implies that
\begin{equation}
    (-X_{j:j+r-2}+X_{j+1:j+r-1})\ket{\psi(A_\alpha)} = (-Y_{\alpha,j:j+r-2}+Y_{\alpha,j+1:j+r-1})\ket{\psi(A_\alpha)}.
\end{equation}
We have therefore constructed a single non-Hermitian $X$ that reproduces the action of the block-dependent $Y_\alpha$ on all injective components simultaneously.
It follows that
\begin{align}
    &(h_{j:j+r-1}+X_{j:j+r-2}-X_{j+1:j+r-1})\ket{\psi(A)} = \sum_{\alpha=1}^\Lambda (h_{j:j+r-1}+X_{j:j+r-2}-X_{j+1:j+r-1})\ket{\psi(A_\alpha)}\nonumber\\
    =& \sum_{\alpha=1}^\Lambda (h_{j:j+r-1}+Y_{\alpha,j:j+r-2}-Y_{\alpha,j+1:j+r-1})\ket{\psi(A_\alpha)} = 0.
\end{align}
By definition, the corresponding $H$ cannot be type III. 
Since the block-injective MPS encompass homogeneous TI MPS of finite bond dimension, this completes our characterization of generalized parent Hamiltonians within this broad class.
Our result further suggests that realizing genuinely type III Hamiltonians in the context of MPS requires going beyond homogeneous MPS representations, for example by introducing nontrivial boundary matrices, as in the W state and Dicke states~\cite{Floridollinas2025Uniform, Gioia2026Distinct}.
\subsection{From a single block-injective MPS to multiple injective MPS}\label{sec:block_to_multiple}
We have shown that a generalized parent Hamiltonian of a block-injective MPS necessarily has the injective components as eigenstates and is of type I or type II. 
This naturally raises the converse question: how to characterize generalized parent Hamiltonians of a set of injective MPS $\{\ket{\psi(A_\alpha)}\}$?
In particular, we ask whether this problem is equivalent to that for a block-injective MPS.
The problem is formulated as finding an extensively local Hermitian Hamiltonian $H$ satisfying
\begin{equation}
    H\ket{\psi(A_\alpha)}
=
E_\alpha\ket{\psi(A_\alpha)}.
\label{eq:multipleMPS}
\end{equation}
We first consider the degenerate case $E_\alpha = E$, such that the whole subspace spanned by these injective MPS forms an eigenspace of $H$.
A necessary condition is that $H$ has a certain state in this subspace, namely $\ket{\psi(\bigoplus_{\alpha=1}^\Lambda A_\alpha)}$, as an eigenstate. 
Since $\ket{\psi(\bigoplus_{\alpha=1}^\Lambda A_\alpha)}$ is always equivalent to a block-injective form, 
by Theorem~\ref{theorem}, generalized parent Hamiltonians of this artificial state automatically have injective MPS in the set as eigenstates, thereby completely solving the degenerate case.
In the non-degenerate case, much stronger constraints are imposed on the MPS.
For example, consider two components with $E_1\neq E_2$, then the corresponding eigenstates must be exactly orthogonal, i.e., $\bra{\psi(A_1)}\psi(A_2)\rangle = 0$. 
The orthogonality holding for sufficiently large but arbitrary sizes implies that the local MPS subspaces from the two states become locally orthogonal beyond some finite length $\ell$.
Thus, we can construct local Hermitian projectors $Q^{(\ell)}_1$ and $Q^{(\ell)}_2$ onto the two subspaces, which are mutually orthogonal.
We may then separate the energy splitting by defining
\begin{equation}
    \tilde{H} = H - \sum\nolimits_j\left(\frac{E_1}{L}Q^{(\ell)}_{1, [j]} + \frac{E_2}{L}Q^{(\ell)}_{2, [j]}\right).
\end{equation}
The two target MPS are then degenerate eigenstates of $\tilde H$, reducing the remaining problem to the degenerate case discussed above.
Consequently, $\tilde H$ can only contain type I or type II contributions.
Meanwhile, the energy-splitting local terms assign target states different eigenvalues, and therefore should be regarded as type I Hamiltonians.
Hence, generalized parent Hamiltonians for multiple matrix product states, i.e., those that satisfy Eq.~(\ref{eq:multipleMPS}), can only be type I or type II.
\section{Solutions of the local tensor equation: Block-injective MPS}\label{sec:block_inj_sol}
\subsection{The commutant algebra and counting formula}\label{sec:counting_block}
We now turn to the local tensor equation of the block-injective MPS.
Ultimately, similar to the injective case, the kernel of the map $X\mapsto T^{(r-1)}_X$ has been shown to be redundant in Sec.~\ref{sec:equivalence_and_redundancy_binj}.
The crucial difference is that, for $r\ge l_{\mathrm{binj}}+1$, this map is no longer surjective onto the full matrix algebra in the doubled virtual space, but only onto the corresponding block-diagonal matrix algebra.
Consequently, not every element of the commutant can be realized as $T_X^{(r-1)}$. The relevant solution space is instead determined by the intersection between the image of the map and the commutant algebra.
The transfer matrix inherits the block structure of the MPS tensor:
\begin{equation}   T_\mathbb{I}^{(1)}=\bigoplus_{\alpha\beta}T_{\mathbb{I},\alpha\beta}^{(1)} =\bigoplus_{\alpha\beta} \sum_{a=1}^{q} A_\alpha^{[a]}\otimes \bar{A}_\beta^{[a]}.
\end{equation}
Correspondingly, every element $\mathcal{E} = T_X^{(r-1)}$ in the image takes the same block-diagonal form:
\begin{equation}
    \mathcal{E}=\bigoplus_{\alpha\beta} \mathcal{E}_{\alpha\beta}.
\label{eq:transferblocknotation}
\end{equation}
Although the full commutant of $T_{\mathbb I}^{(1)}$ may contain elements coupling different $(\alpha\beta)$-blocks when the spectra of block-resolved transfer matrices overlap, such off-diagonal elements do not belong to the image of the map and are therefore irrelevant to the counting.
Restricting to the block-diagonal algebra, the commuting condition of Eq.~(\ref{S_eq:transfer_matrix_comm}) factorizes into independent conditions in each block,
\begin{equation}
    [\mathcal{E}_{\alpha\beta}, T_{\mathbb{I},\alpha\beta}^{(1)}] = 0, \quad\forall\alpha,\beta.
\end{equation}
Hence, the total number of independent solutions is obtained by summing the contribution from all $(\alpha\beta)$-block transfer matrices $T_{\mathbb I,\alpha\beta}^{(1)}$. 
We next characterize the redundant solutions identified in
Sec.~\ref{sec:equivalence_and_redundancy_binj} in terms of $\delta X$.
In that section, we have shown that
$Y_{1:\tilde r-1}$ acts as a scalar on the local MPS subspace of each injective component: $Y_{1:\tilde r-1}
    \ket{A_{\alpha,\mu\nu}^{(\tilde r-1)}}
    =
    c_\alpha\ket{A_{\alpha,\mu\nu}^{(\tilde r-1)}}$,
where the redundant solution $\delta X$ is the anti-Hermitian part of $Y$.
Similar to Eq.~\eqref{eq:injective_redudant}, we therefore obtain
\begin{align}\label{eq:block_injective_redundant}
    T_{\delta X,\alpha\beta}^{(r-1)}
    \left(T_{\mathbb I,\alpha\beta}^{(1)}\right)^{\tilde r-r}
    =
    \left(
    \frac{c_{\alpha,R}-c_{\beta,R}}{2}
    +i\frac{c_{\alpha,I}+c_{\beta,I}}{2}
    \right)
    \left(T_{\mathbb{I},\alpha\beta}^{(1)}\right)^{\tilde r-1}.
\end{align}
We apply a similar strategy as in the injective case by decomposing $\delta X$ into a particular solution and a remaining part.
First we determine the dimension of what we call particular redundant solutions, which are analogous to the $\delta X = i\ \text{Im}(c)\mathbb{I}$ in the injective case (see Sec.~\ref{sec:equivalence_and_redundancy_inj}).
For $r\ge l_\mathrm{binj}+1$, the generalized transfer matrix is surjective onto the diagonal blocks of the doubled virtual space.
Consequently, this dimension is given by the number of independent real coefficients on the right-hand side of Eq.~\eqref{eq:block_injective_redundant}.
For an injective MPS, it reduces to a single parameter $c_I$ [Eq.~\eqref{eq:injective_redudant}].
Here, for a block-injective MPS, the $(\alpha,\alpha)$-blocks only depend on the imaginary parts $c_{\alpha,I}$, yielding $\Lambda$ real parameters.
For the blocks with $\alpha\neq\beta$, the real parts $\{c_{\alpha, R}\}$ can enter in the form $c_{\alpha, R}-c_{\beta, R}$.
If all the block-resolved transfer matrices $T_{\mathbb I,\alpha\beta}^{(1)}$ are nonzero, these blocks contribute $\Lambda-1$ additional independent real coefficients.
However, some $T_{\mathbb I,\alpha\beta}^{(1)}$ may vanish (as in the GHZ state), or more generally vanish upon grouping multiple sites, which could change the number of particular solutions.
To account for this, we note that $\left(T_{\mathbb I,\alpha\beta}^{(1)}\right)^r = 0$ implies that the $r$-local MPS subspaces of $A_\alpha$ and $A_\beta$ are mutually orthogonal.
We therefore consider the finest partition of the $\Lambda$ injective blocks into $k$ sets, $\mathcal{A}_1,\cdots, \mathcal{A}_k$, containing $\Lambda_1,\ldots,\Lambda_k$ blocks, respectively, with the condition that the $r$-local MPS subspaces associated with blocks belonging to different sets are mutually orthogonal.
Here, ``finest'' means that none of the sets can be further decomposed under the requirement.
Therefore, the total number of particular redundant solutions, accounting for contributions from both imaginary and real parts of the coefficients, are
\begin{equation}
    \Lambda+(\Lambda_1-1)+(\Lambda_2-1)+\cdots+(\Lambda_k-1) = 2\Lambda-k.
\end{equation}
As in the injective case, the remaining redundant solutions, denoted by $\delta X'$, satisfy
\begin{equation}
    T_{\delta X',\alpha\beta}^{(r-1)}
    \left(T_{\mathbb I,\alpha\beta}^{(1)}\right)^{\tilde r-r}
    =0.
\end{equation}
Again, by block-injectivity, this condition can be analyzed independently in each $(\alpha,\beta)$ sector.
As in the injective case, the redundant solutions are associated with the zero-eigenspace and zero-eigenvalue Jordan blocks of the block-resolved transfer matrix, which are easy to count using similar methods.
Having determined the dimension of the redundant solution space, we can now obtain the number of inequivalent type II Hamiltonians.
When all block-resolved transfer matrices are diagonalizable, the result is
\begin{equation}\label{S_eq:counting_formula_binj}
    \mathcal{N} = (\sum_{\alpha\beta}\sum_{\tau_{\lambda,\alpha\beta}\neq0}d_{\lambda, \alpha\beta}^2)-2\Lambda + k,
\end{equation}
which proves Eq.~(16) in the main text.
The non-diagonalizable case follows analogously by applying Eq.~\eqref{S_eq:counting_formula_inj_Jordan} to each block-resolved transfer matrix.

\subsection{Action of a type II Hamiltonian on injective components}\label{sec:individual_action}
In this section, we study how a type II Hamiltonian acts on the individual injective components. 
A natural basis of solutions of the local tensor equation can be chosen such that $T_X^{(r-1)}$ is supported within each block.
For a fixed $(\alpha\beta)$-block, the corresponding solution can be chosen to satisfy
\begin{equation}
    [T_{X, \alpha\beta}^{(r-1)}, T_{\mathbb{I}, \alpha\beta}^{(1)}] = 0,\qquad T_{X, \alpha'\beta'}^{(r-1)} = 0,\quad \forall (\alpha'\beta')\neq(\alpha\beta).
\end{equation}
The existence of a solution is guaranteed by block-injectivity.
These solutions naturally separate into two cases, which show that the generalized parent Hamiltonian can be (i) type II on some of the injective components, or (ii) type I on each of its injective components yet type II for the full block-injective MPS.
\begin{itemize}
\item[(i)] $\alpha=\beta$.
In this case,
\begin{equation}
     [T_{X, \alpha\alpha}^{(r-1)}, T_{\mathbb{I}, \alpha\alpha}^{(1)}] = 0,\qquad T_{X, \beta\beta}^{(r-1)} = 0,\quad \forall \beta\neq\alpha.
\end{equation}
Hence, on the injective component $\ket{\psi(A_\alpha)}$, the solution obeys the same local tensor equation as in the injective case and can give rise to a type II Hamiltonian. On every other injective component, $X$ is in the kernel of the corresponding map, so the solution is redundant, i.e., type I. 

\item[(ii)] $\alpha\neq\beta$.
It follows that
\begin{equation}
    T_{X, \alpha\alpha}^{(r-1)} = 0,\quad \forall \alpha.
\end{equation}
Therefore, when restricted to any individual injective component $\ket{\psi(A_\alpha)}$, the solution is type I.
As before, the remaining equations serve as consistency conditions. 
Note that there is no contradiction between being type I on individual injective components and type II on the block-injective MPS.
Global type I requires the existence of a Hermitian local decomposition of the Hamiltonian, whose local terms annihilate all of the injective components simultaneously. Here, the tensor equations do not imply such a Hermitian decomposition.
\end{itemize}

\section{Type II Hamiltonians for the Majumdar-Ghosh ground states}

In this section, we apply the construction of type II Hamiltonians developed for block-injective MPS to the two degenerate ground states of the Majumdar-Ghosh model~\cite{Majumdar1969NNN}. 
These are two dimer states defined on  a spin-$1/2$ chain of $2L$ sites under PBC, which are products of nearest-neighbor spin singlets with different pairing patterns:
\begin{align}
    \ket{\psi_\mathrm{o}}  &= \bigotimes_{j=1}^L \frac{1}{\sqrt{2}}\left(\ket{\uparrow}_{2j}\ket{\downarrow}_{2j+1} - \ket{\downarrow}_{2j}\ket{\uparrow}_{2j+1}\right),\nonumber\\
    \ket{\psi_\mathrm{e}}  &= \bigotimes_{j=1}^L \frac{1}{\sqrt{2}}\left(\ket{\uparrow}_{2j-1}\ket{\downarrow}_{2j} - \ket{\downarrow}_{2j-1}\ket{\uparrow}_{2j}\right),
\end{align}
where $\mathrm{o}$ and $\mathrm{e}$ are shorthand for odd and even, respectively. 
Denoting the spin-$1/2$ operators by $s_j^a = \sigma_j^a/2, ~ a = x, y, z$, the Majumdar-Ghosh Hamiltonian can be written as a sum of three-site operators $(\vec{s}_{j-1} + \vec{s}_{j} + \vec{s}_{j+1})^2$, of which the minimal value is attained when the three-site total spin equals $1/2$, as in the two dimer states.
This shows that the Majumdar-Ghosh Hamiltonian is a frustration-free (type I parent) Hamiltonian for the two dimer states.
We aim to characterize the type II Hamiltonians of the set of two dimer states, i.e., $\{\ket{\psi_\mathrm{o}}, \ket{\psi_\mathrm{e}}\}$.
As discussed in Sec.~\ref{sec:block_to_multiple}, since the two states are non-orthogonal, this problem is equivalent to constructing type II Hamiltonians for the block-injective MPS representing their superposition, $\ket{\psi} = \ket{\psi_\mathrm{o}} + \ket{\psi_\mathrm{e}}$. 
To apply the method developed in Sec.~\ref{sec:block_inj_sol}, we first express both dimer states as injective TI MPS.
To this end, we group each pair of neighboring sites $2j-1$ and $2j$ into a single site $j$ with Hilbert space dimension $q=4$.
As implied by the spin fusion rule $\mathbf{\frac{1}{2}}\otimes \mathbf{\frac{1}{2}} = \mathbf 0\oplus \mathbf 1$, we choose a local basis consisting of the spin singlet and a Cartesian basis for the spin triplets:
\begin{align}
&\ket{0}_j
=\frac{1}{\sqrt{2}}\left(\ket{\uparrow}_{2j-1}\ket{\downarrow}_{2j} - \ket{\downarrow}_{2j-1}\ket{\uparrow}_{2j}\right)
,
&\ket{z}_j
=
\frac{1}{\sqrt{2}}\left(\ket{\uparrow}_{2j-1}\ket{\downarrow}_{2j} + \ket{\downarrow}_{2j-1}\ket{\uparrow}_{2j}\right)\nonumber\\
&\ket{x}_j
=
\frac{1}{\sqrt{2}}\left(\ket{\downarrow}_{2j-1}\ket{\downarrow}_{2j} - \ket{\uparrow}_{2j-1}\ket{\uparrow}_{2j}\right),
&\ket{y}_j
=
\frac{i}{\sqrt{2}}\left(\ket{\downarrow}_{2j-1}\ket{\downarrow}_{2j} + \ket{\uparrow}_{2j-1}\ket{\uparrow}_{2j}\right).
\end{align}
In this basis, the state $\ket{\psi_\mathrm{o}}$ admits an injective MPS representation with bond dimension $\chi_\mathrm{o}=2$:
\begin{equation}
    A_\mathrm{o}^{[0]} = -\frac{1}{2}I_2,\qquad A_\mathrm{o}^{[a]} = \frac{1}{2}\sigma^a, a=x,y,z.
\end{equation}
The state $\ket{\psi_\mathrm{e}}$ becomes a product state, and hence admits a representation with $\chi_\mathrm{e}=1$:
\begin{equation}
    A_\mathrm{e}^{[0]} = 1,\qquad A_\mathrm{e}^{[a]} =0, a=x,y,z.
\end{equation}
For the superposition state $\ket{\psi}$, the local tensor $A = A_\mathrm{o}\oplus A_\mathrm{e}$ is a block-injective MPS representation with a block-injectivity length $l_\mathrm{binj} = 2$.
We therefore seek solutions to $T_X^{(2)} = \mathcal{E}$, where $\mathcal{E}$ runs over block-diagonal elements of the commutant of the transfer matrix $T_\mathbb{I}^{(1)}$. 
The transfer matrix takes the form
\begin{equation}
    T_\mathbb{I}^{(1)} = |\mathbb{I})\!)(\!(\mathbb{I}|\oplus (-\frac{1}{2}I_2)\oplus (-\frac{1}{2}I_2)\oplus 1,
\label{eq:MGTmatrix}
\end{equation}
where the four blocks are ordered as $(\mathrm{oo})$, $(\mathrm{oe})$, $(\mathrm{eo})$, and $(\mathrm{ee})$, and the state in the doubled virtual space $|\mathbb{I})\!)$ follows the definition in Eq.~\eqref{eq:vectorized_operator}.
Thus, $(\mathrm{oo})$- and $(\mathrm{ee})$-block transfer matrices each have a single nonzero eigenvalue, while the $(\mathrm{oe})$- and $(\mathrm{eo})$-block transfer matrices each have a doubly degenerate eigenvalue.
Finally, since the two injective MPS are non-orthogonal, we have $k=1$.
Applying Eq.~(\ref{S_eq:counting_formula_binj}), we obtain the number of inequivalent type II Hamiltonians $\mathcal{N} = 7$.
Notably, both $(\mathrm{oo})$- and $(\mathrm{ee})$-block transfer matrices have rank one.
Consequently, the two dimer states are zero correlation length states, as is also clear from their product-state structure in the underlying spin-$1/2$ representation, and thus exhibit no type II Hamiltonian individually. 
This corresponds to the case discussed in Sec.~\ref{sec:individual_action}, where the type II Hamiltonians admit no Hermitian decomposition satisfying the type I condition for both states simultaneously.
Although we solve the transfer matrix equation for the anti-Hermitian operator $X$ represented in the grouped $q=4$ Hilbert space, we will express the corresponding type II Hamiltonians in terms of the original spin-$1/2$ operators, which provides a more transparent physical interpretation.
We now present the seven inequivalent type II Hamiltonians obtained from the transfer matrix equation.
Similar to the AKLT case in Sec.~\ref{sec:AKLT}, we organize the basis elements of the commutant according to their transformation properties under $\mathrm{SU}(2)$ symmetry, which separates them into three categories: one-dimensional scalar and two distinct three-dimensional vectors.
As we show below, the corresponding type II Hamiltonians inherit the same transformation properties.
For each category, we solve for $X$, present the local Hermitian terms, and explicitly verify that they satisfy the required local annihilation condition.
\subsection{Scalar spin chirality}
Among the block-diagonal elements in the commutant, there are four scalars:
\begin{align}
    \mathcal{E}_\mathrm{o} &= i |\mathbb{I})\!)(\!(\mathbb{I}|\oplus 0 \oplus 0 \oplus 0, &\mathcal{E}_\mathrm{e} &= 0\oplus 0 \oplus 0 \oplus i,\nonumber\\
    \mathcal{E}_{R} &= 0\oplus I_2 \oplus -I_2 \oplus 0 ,
    &\mathcal{E}_{I} &= 0\oplus iI_2 \oplus iI_2 \oplus 0,
\end{align}
where we have used the same block ordering as in Eq.~(\ref{eq:MGTmatrix}).
These operators are invariant under $\mathrm{SU}(2)$ rotations on the doubled virtual space.
Only one direction in the linear subspace spanned by these operators is non-redundant.
To see this, we apply the equation of redundant solutions \eqref{eq:block_injective_redundant} with two complex coefficients $c_\mathrm{o}$ and $c_\mathrm{e}$:
\begin{align}
    T_X^{(2)} = \mathcal{E} =& ic_{\mathrm{o}, I} |\mathbb{I})\!)(\!(\mathbb{I}|
    \oplus  
    \frac{1}{4}\left(
    \frac{c_{\mathrm{o},R}-c_{\mathrm{e},R}}{2}
    +i\frac{c_{\mathrm{o},I}+c_{\mathrm{e},I}}{2}
    \right)I_2
    \oplus 
    \frac{1}{4}\left(
    \frac{c_{\mathrm{e},R}-c_{\mathrm{o},R}}{2}
    +i\frac{c_{\mathrm{o},I}+c_{\mathrm{e},I}}{2}
    \right)I_2
    \oplus
    ic_{\mathrm{e}, I}\nonumber\\
    =& c_{\mathrm{o}, I}\mathcal{E}_\mathrm{o} + c_{\mathrm{e}, I}\mathcal{E}_\mathrm{e} + \frac{c_{\mathrm{o},R}-c_{\mathrm{e},R}}{8}\mathcal{E}_{R} + \frac{c_{\mathrm{o},I}+c_{\mathrm{e},I}}{8}\mathcal{E}_{I},
\end{align}
where $c_{\alpha,R}$ and $c_{\alpha,I}$ denote the real and imaginary parts of $c_\alpha$, respectively.
The formula shows that all the $(2\Lambda-k = 3)$-dimensional particular redundant solutions are supported in this four-dimensional scalar sector.
We therefore choose one representative scalar, $\mathcal{E}_\mathrm{o}$, as an inequivalent solution.
An explicit anti-Hermitian solution satisfying $T_X^{(2)} = \mathcal{E}_\mathrm{o}$ can be chosen to be one-site: 
\begin{equation}
    X_{j} = i\sum_{a=x,y,z}
\left(\ket{a}\bra{a}\right)_{j} =  i (\mathbb{I}-\ket{0}\bra{0})_{j}.
\end{equation}
In terms of the underlying spin-$1/2$ operators, this can be expressed as $X_j=i(\vec{s}_{2j-1}\cdot\vec{s}_{2j}+3/4)$.
To construct the corresponding type II Hamiltonian, we calculate the action of the anti-Hermitian telescoping term on the $2$-local MPS:
\begin{align}
    (X_j-X_{j+1})\ket{A^{(2)}}_{j,j+1} = &i(\vec{s}_{2j-1}\cdot \vec{s}_{2j}-\vec{s}_{2j+1}\cdot \vec{s}_{2j+2})\ket{A^{(2)}}_{j,j+1}\nonumber\\
    =& -i[\vec{s}_{2j-1}\cdot \vec{s}_{2j}-\vec{s}_{2j+1}\cdot \vec{s}_{2j+2}, \vec{s}_{2j}\cdot \vec{s}_{2j+1}]\ket{A^{(2)}}_{j,j+1}\nonumber\\
    =& \left(-\vec{s}_{2j-1}\cdot(\vec{s}_{2j}\times \vec{s}_{2j+1})-\vec{s}_{2j}\cdot(\vec{s}_{2j+1}\times \vec{s}_{2j+2})\right)\ket{A^{(2)}}_{j,j+1}.
\end{align}
Here, the second equality follows from the local singlet structure of the two dimer states, and the last equality follows directly from the spin-$1/2$ commutation relations.

We therefore obtain that the corresponding type II Hamiltonian is the scalar spin chirality:
\begin{equation}
    H = \sum\nolimits_j C_{j,j+1,j+2} = \sum\nolimits_j \vec{s}_j\cdot(\vec{s}_{j+1}\times \vec{s}_{j+2}).
\end{equation}
Analogous to the scalar solution for the AKLT ground state (Sec.~\ref{sec:AKLT_scalar}), this Hamiltonian has previously also been identified as a generalized parent Hamiltonian of the two dimer states~\cite{Sanada2023Quantum}, while the type II-ness has not been established there.

\subsection{Total spin operators}
We next consider the three Hermitian elements of the commutant supported in $\mathrm{oe}$ and $\mathrm{eo}$ blocks:
\begin{equation}
    \mathcal{E}_a = 0\oplus \sigma^a \oplus (-\sigma^a)^* \oplus 0
    \qquad a=x,y,z.
\end{equation}
The three elements lead to one-site anti-Hermitian \(X\) operators parameterized by a real-valued vector $\vec{v}$:
\begin{equation}
    X_{j}
    =   
    \ket{0}_{j}\bra{\vec{v}}_{j}
    -
    \ket{\vec{v}}_{j}\bra{0}_{j},
    \qquad \ket{\vec{v}}_{j} = v_x\ket{x}_{j} + v_y\ket{y}_{j}+v_z\ket{z}_{j}.
\end{equation}
To identify the corresponding type II Hamiltonians, we evaluate the action of the telescoping term on the $2$-local MPS.
The action on $\ket{\psi_\mathrm{e}}$ reads:
\begin{align}
   (X_j-X_{j+1})\ket{A_\mathrm{e}^{(2)}}_{j,j+1} = & \left( \vec v\cdot  (\vec{s}_{2j-1}-\vec{s}_{2j})\mathcal{P}_{2j-1,2j} -  \vec v\cdot  (\vec{s}_{2j+1}-\vec{s}_{2j+2})\mathcal{P}_{2j+1,2j+2}
   \right)\ket{A_\mathrm{e}^{(2)}}_{j,j+1}\nonumber\\
   = &2 \vec v\cdot (\vec s_{2j} + \vec s_{2j+1})\ket{A_\mathrm{e}^{(2)}}_{j,j+1},
\end{align}
where $\mathcal{P}_{i,j}$ denotes the permutation operator exchanging the spin-$1/2$'s at sites $i$ and $j$, and the second equality follows from the local singlet structure of the dimer state.
On the other hand, for $\ket{\psi_\mathrm{o}}$ there is
\begin{equation}
    2 \vec v\cdot (\vec s_{2j} + \vec s_{2j+1})\ket{A_\mathrm{o}^{(2)}}_{j,j+1} = 0.
\end{equation}

The two relations allow us to construct a common telescoping operator for the block-injective MPS. Applying Eq.~\eqref{eq:block_telescoping}, we introduce the non-Hermitian operator
\begin{equation}
    \widetilde{X}_{j,j+1} = 2X_j(\Pi_{\mathrm{e}}^{(2)})_{j, j+1}.
\end{equation}
It follows that the anti-Hermitian part of $\widetilde X$ satisfies $T^{(2)}_{\frac{1}{2}(\tilde{X}-\widetilde{X}^\dagger)} = T^{(2)}_X$, implying that $\widetilde{X}$ and $X$ are equivalent telescoping terms, with their difference corresponding to a redundant solution.
We therefore obtain the local annihilation relation
\begin{equation}
   \left( -4 \vec v\cdot (\vec s_{2j} + \vec s_{2j+1}) + \widetilde{X}_{j,j+1} - \widetilde{X}_{j+1,j+2}\right)\ket{A^{(3)}}_{j,j+1,j+2} = 0,
\end{equation}
and obtain that the corresponding type II Hamiltonians are the three components of the total spin $\sum_j{s^\alpha_j}$ for $\alpha \in \{x,y,z\}$, which form a vector under $\mathrm{SU}(2)$ rotation.

The fact that the total spin annihilates both dimer states is rather straightforward: each state is a product of spin singlets, and each singlet is annihilated by total spin operators acting on its two sites~\cite{Shiraishi2017Systematic}.
Our construction establishes the correspondence between these generalized parent Hamiltonians and certain solutions $X$ of the transfer matrix equation, thereby identifying them as a three-dimensional class of type II Hamiltonians, similar to the case of the AKLT state.
\subsection{Staggered vector spin chirality}
The remaining three elements in the commutant read
\begin{equation}
    \mathcal{E}_a = 0\oplus i\sigma^a \oplus i(\sigma^a)^* \oplus 0
    \qquad a=x,y,z.
\end{equation}
We construct the corresponding anti-Hermitian $X$ operator as
\begin{equation}
    X_{j,j+1} = i\left(\ket{0}\bra{\vec{v}}+\ket{\vec{v}}\bra{0}\right)_j (\Pi_{\mathrm{e}}^{(2)})_{j, j+1}+(\Pi_{\mathrm{e}}^{(2)})^\dagger_{j, j+1}i\left(\ket{0}\bra{\vec{v}}+\ket{\vec{v}}\bra{0}\right)_j
\end{equation}
The key observation is that the one-site operator in this equation produces the desired generalized transfer matrices in $\mathrm{ee}$, $\mathrm{oe}$ and $\mathrm{eo}$ blocks, whereas the two-site operator $\Pi_{\mathrm{e}}^{(2)}$ is introduced to eliminate unwanted contributions in the $\mathrm{oo}$ block. 

To construct the corresponding type II Hamiltonians, it is convenient to introduce the non-Hermitian operator
\begin{equation}
    \widetilde X_{j,j+1} = 2i\left(\ket{0}\bra{\vec{v}}+\ket{\vec{v}}\bra{0}\right)_j (\Pi_{\mathrm{e}}^{(2)})_{j, j+1},
\end{equation}
whose anti-Hermitian part is precisely \(X\).
Therefore, replacing \(X\) by \(\widetilde X\) amounts to adding a Hermitian telescoping term which leaves the global Hamiltonian unchanged.

We now evaluate the action of the non-Hermitian telescoping term on the local MPS:
\begin{align}
   (\widetilde X_{j,j+1}-\widetilde X_{j+1, j+2})\ket{A^{(3)}}_{j,j+1,j+2} = & 2i\left( \vec v\cdot  (\vec{s}_{2j-1}-\vec{s}_{2j}) -  \vec v\cdot  (\vec{s}_{2j+1}-\vec{s}_{2j+2})
   \right)\ket{A^{(3)}}_{j,j+1,j+2}\nonumber\\
   = &-4 \vec v\cdot  (\vec{s}_{2j-1} + \vec s_{2j+2})\times(\vec{s}_{2j}+\vec{s}_{2j+1}) 
  \ket{A^{(3)}}_{j,j+1,j+2}.
\end{align}
From the local condition, we can extract the corresponding type II Hamiltonian
\begin{equation}\label{eq:stagger}
    H = \sum_{j=1}^L \vec v\cdot  (\vec{s}_{2j-1} + \vec s_{2j+2})\times(\vec{s}_{2j}+\vec{s}_{2j+1}) = \sum_{j=1}^{2L} (-1)^{j+1}~\vec v\cdot (\vec{s}_{j}\times\vec{s}_{j+2}).
\end{equation}
This Hamiltonian describes staggered vector spin chirality between next-nearest-neighbor spins, and transform as a vector under $\mathrm{SU}(2)$ rotation.

Again, the structure of dimer states makes it straightforward to understand how Eq.~\eqref{eq:stagger} acts as a generalized parent Hamiltonian: each vector chirality term $\vec{s}_j\times\vec{s}_{j+2}$ has the same action as its one-site translation to the left or right, depending on the pairing pattern. The alternating signs in Eq.~\eqref{eq:stagger} then ensure that these contributions cancel pairwise.

\section{Parent Hamiltonians via MPS Commutators}

In this section, we present a complementary perspective on the construction of some parent Hamiltonians,  which may also be of independent interest.
The basic idea is to require that the action of the local Hermitian term $h_j$ of the generalized parent Hamiltonian (which can also be longer-ranged) can be pushed through the MPS tensor as a commutator in the virtual space:
\begin{equation}\label{Seq:push_through}
    \sum_{b=1}^q h^a_{b} A^{[b]} = [C, A^{[a]}],
\end{equation}
where $C$ is a matrix of dimension $\chi$. 
Upon summing the local terms $h_j$ over the 1D chain, the commutator contributions telescope in the virtual space, thereby ensuring that the MPS is an eigenstate of the global Hamiltonian.
To construct such local Hermitian terms, we therefore search over matrices $C$ for which the above relation gives rise to a Hermitian local operator $h$.
More generally, such a relation may arise only after blocking several physical sites into the local MPS tensor, or when considering local Hermitian terms with a longer range. This kind of construction has previously appeared in studies of generalized parent Hamiltonians for the PXP model and related models~\cite{lin2019exact, Ivanov2025Many}.
Importantly, the push-through relation provides a sufficient, but not necessary, condition for generalized parent Hamiltonians, as implied by Ref.~\cite{Rubio2026Local}. 
It therefore generally captures a subclass of type II Hamiltonians, in contrast to the exhaustive construction presented in the main text.

In the following, we focus on the single-site case, for which the generalization to the multi-site case is straightforward.
We impose an additional condition that facilitates a systematic construction of $h$: the set of local tensors $\{A^{[a]}\}$ forms an orthonormal basis of its linear span $W$ up to a normalization factor $\lambda$:
\begin{equation}
    \operatorname{Tr}\left[A^{[a]\dagger} A^{[b]} \right]= \lambda \delta_{ab}.
\end{equation}
For Eq.~\eqref{Seq:push_through} to admit a solution, the commutators $[C, A^{a}]$ must remain in the same subspace $W$.
One condition that automatically guarantees this property is single-site injectivity, for which $W$ coincides with the full matrix algebra $M_{\chi}(\mathbb{C})$ in the virtual linear space. 
A weaker condition, which we refer to as \textit{ideality}, requires $W$ to be left invariant under the adjoint action of $C$, $\mathrm{Ad}_C(\cdot)=[C,\cdot]$:
\begin{equation}
    \mathrm{Ad}_C (W) \subseteq W, \quad \forall C\in M_{\chi}(\mathbb{C}).
\end{equation}
In this case, $W$ forms a Lie ideal of $M_{\chi}(\mathbb{C})$ with respect to the commutator.
Furthermore, a closely related construction applies when $W$ is a Lie subalgebra of $M_{\chi}(\mathbb{C})$, with the additional constraint that the matrix $C$ cannot be chosen arbitrarily but must belong to the same subalgebra.

Given the orthogonality and ideality conditions above, we can solve Eq.~\eqref{Seq:push_through} by construction. 
Ideality ensures that the commutator $[C, A^{[a]}]$ remains within $W$, while orthogonality allows its expansion coefficients to be extracted by projection onto the basis elements.
Specifically, the matrix elements of $h$ are given by:
\begin{equation}\label{Seq:matrix_elements}
    h_b^a = \operatorname{Tr}\left(A^{[b]\dagger}[C, A^{[a]}]\right)/\lambda.
\end{equation}
When $C$ is Hermitian, these matrix elements satisfy $h_b^a = (h_a^b)^*$, and therefore $h$ is automatically Hermitian as desired. Excluding the zero solution from $C = \mathds{1}$, the linear space of solutions has dimension at most $\chi^2-1$. In the following, we discuss two examples; however, note that this method by itself does not guarantee that the resulting parent Hamiltonian is type II

\textit{\textbf{AKLT ground state---}}We first revisit the AKLT ground state. Previously, we have identified nine type II Hamiltonians with a range up to three sites. The method developed here allows us to investigate the one-site solutions in greater detail.
The local MPS tensors are
\begin{equation}
    \begin{split}
        A^{[+1]}= \sqrt{\frac{2}{3}} \hat\sigma^+, \quad A^{[0]} = -\frac{1}{\sqrt{3}} \hat\sigma^z, \quad A^{[-1]} = -\sqrt{\frac{2}{3}} \hat\sigma^- ,
    \end{split}
\end{equation}
which form an equal-norm orthogonal basis of $W = \mathrm{sl}_2(\mathbb{C})$, the space of traceless $2\times 2$ complex matrices, which is a Lie ideal under the commutator. We then choose $C= \sigma^{\alpha}, \alpha = x, y,z$, and construct the corresponding local Hermitian terms $h$ via Eq.~\eqref{Seq:matrix_elements}. 
The resulting operators are precisely the three spin-$1$ operators $S^{\alpha}$, constituting the single-site solution space from the push-through construction.
We should remark that the systematic construction explained in this section works only if the basis is orthonormal (up to a normalization factor).
In the case of the AKLT state, instead, the two-site or three-site local tensors do not satisfy the orthonormality condition, thus not allowing for a straightforward derivation of local Hermitian terms that push through as commutators in virtual space.
Moreover, solving the push-through relation for two-site Hamiltonians can lead to the two-site extension of one-site solutions, which is not straightforward to exclude systematically.

\textit{\textbf{Cluster state---}}The cluster state admits an MPS representation with local tensors
\begin{equation}
    A^{[\uparrow]}=\begin{pmatrix}
1 & 1\\
0 & 0 
\end{pmatrix}, \qquad A^{[\downarrow]} = \begin{pmatrix}
0 & 0\\
1 & -1 
\end{pmatrix}.
\end{equation}
After blocking two neighboring sites, the resulting set of local MPS tensors forms an orthogonal basis of the full matrix space and is therefore injective, as can be seen from 
\begin{equation}
    \begin{split}
        & A^{[\uparrow]}A^{[\uparrow]}=\begin{pmatrix}
1 & 1\\
0 & 0 
\end{pmatrix}, \;\;\; A^{[\downarrow]}A^{[\downarrow]}=\begin{pmatrix}
0 & 0\\
-1 & 1 
\end{pmatrix}, \\
        & A^{[\uparrow]}A^{[\downarrow]}=\begin{pmatrix}
1 & -1\\
0 & 0 
\end{pmatrix}, \;\;\;A^{[\downarrow]}A^{[\uparrow]}=\begin{pmatrix}
0 & 0\\
1 & 1 
\end{pmatrix},
    \end{split}
\end{equation}
We therefore choose $C= \sigma^{\alpha}, \alpha = x, y,z$ and construct the corresponding two-site Hermitian operators:
\begin{equation}\label{eq:cluster_commutator}
     h_{j,j+1}^{(\sigma^x)} = (\sigma_j^x-1) \sigma_{j+1}^z,\;\;\;
     h_{j,j+1}^{(\sigma^y)} = \sigma_j^z  \sigma_{j+1}^y-\sigma_j^y  \sigma_{j+1}^z,\;\;
     h_{j,j+1}^{(\sigma^z)} = \sigma_j^z (1-\sigma_{j+1}^x).
\end{equation}
However, note that in this case these are not truly type II operators.
The transfer matrix equation developed in this work provides an alternative approach to characterizing the equivalence classes of type II Hamiltonians.
Since the transfer matrix of the cluster state has a single nonzero eigenvalue, no type II Hamiltonians exist, and therefore all the generalized parent Hamiltonians obtained from the commutator construction above must be of type I.
We now verify this conclusion explicitly by adding Hermitian telescoping terms to the local Hamiltonians in Eq.~\eqref{eq:cluster_commutator} such that they annihilate the cluster state locally.
The key observation is that the cluster state is a stabilizer state satisfying $K_j\ket{\psi} = \ket{\psi}$ for Pauli strings $K_j=\sigma_j^z\sigma_{j+1}^x\sigma_{j+2}^z$. Using these stabilizer relations, the action of $h_{j,j+1}^{(\sigma^x)}$ can be rewritten as
\begin{equation}
    h_{j,j+1}^{(\sigma^x)}\ket{\psi} = (\sigma_{j-1}^zK_{j-1}- \sigma_{j+1}^z)\ket{\psi} = (\sigma_{j-1}^z- \sigma_{j+1}^z)\ket{\psi}.
\end{equation}
It follows that the Hermitian local operator $(h_{j,j+1}^{(\sigma^x)}- \sigma_{j-1}^z + \sigma_{j+1}^z)$ has the cluster state as an eigenstate, verifying the type I-ness. Similarly, Hermitian local annihilators from the other two Hamiltonians can be constructed as
\begin{equation}
    h_{j,j+1}^{(\sigma^y)} - \sigma_{j+2}^y  \sigma_{j+3}^z+\sigma_j^y  \sigma_{j+1}^z, \qquad h_{j,j+1}^{(\sigma^z)}- \sigma_{j}^z + \sigma_{j+2}^z.
\end{equation}
This cluster-state example demonstrates that, although the commutator method generates valid generalized parent Hamiltonians, it does not by itself distinguish type II from type I ones.

\bibliography{Heran_scar}